\documentclass[fleqn,proof]{pasj01}
\usepackage{bm}
\usepackage{subfigure}

\Received{$\langle$2020 March 10 $\rangle$}
\Accepted{$\langle$2020 June 19$\rangle$}
\Published{$\langle$publication date$\rangle$}

\begin{document}

\title{Numerical simulation of solar photospheric jet-like phenomena caused by magnetic reconnection}
\author{Yuji K{\small OTANI}\altaffilmark{1} and Kazunari S{\small HIBATA}\altaffilmark{2}}%
\altaffiltext{1}{Department of Astronomy, Graduate School of Science, Kyoto University, Kitashirakawa-Oiwake-cho, Sakyo-ku, Kyoto, Kyoto 606-8502, Japan}
\altaffiltext{2}{Kwasan and Hida Observatories, Kyoto University, 17 Ohmine-cho Kita Kazan, Yamashina-ku, Kyoto, Kyoto 607-8471, Japan}
\email{kotani@kusastro.kyoto-u.ac.jp}


\maketitle

\begin{abstract}
Jet phenomena with a bright loop in their footpoint, called anemone jets, have been observed in the solar corona and chromosphere. 
These jets are formed as a consequence of magnetic reconnection, and from the scale universality of magnetohydrodynamics (MHD), it can be expected that anemone jets exist even in the solar photosphere.
However, it is not necessarily apparent that jets can be generated as a result of magnetic reconnection in the photosphere, where the magnetic energy is not dominant.
Furthermore, MHD waves generated from the photospheric jets could contribute to chromospheric heating and spicule formation; however, this hypothesis has not yet been thoroughly investigated.
In this study, we perform 3D MHD simulation including gravity with the solar photospheric parameter to investigate anemone jets in the solar photosphere.
In the simulation, jet-like structures were induced by magnetic reconnection in the solar photosphere.
We determined that these jet-like structures were caused by slow shocks formed by the reconnection and were propagated approximately in the direction of the background magnetic field.
We also suggested that MHD waves from the jet-like structures could influence local atmospheric heating and spicule formation. 
\end{abstract}

\section{Introduction}

     In the solar atmosphere, different kinds of jet phenomena, that is, the ejection of collimated plasma, can be observed at different wavelengths.
In particular, numerous jets with a bright loop at their footpoint can be observed.
These jets are called inverted-Y, Eiffel tower-shaped, or anemone jets because of their appearance (e.g., \cite{1994ApJ...431L..51S,2007Sci...318.1591S,2009SoPh..259...87N}).
Herein, we use the term “anemone jets" to refer to these jets in this paper.

Anemone jets were first discovered in the corona by the Yohkoh satellite \citep{1992PASJ...44L.173S,1996PASJ...48..123S}.
\citet{1999Ap&SS.264..129S} proposed that these jets are formed as a consequence of magnetic reconnection similar to large-scale flares followed by coronal mass ejections.
Today, numerous simulations have been performed in two-dimensional (2D) \citep{1996PASJ...48..353Y} and three-dimensional (3D) scenarios (\cite{2008ApJ...673L.211M}, 2013; \cite{2009ApJ...704..485T}; \cite{2009ApJ...691...61P}, 2015a, 2016; \cite{2013ApJ...769L..21A}; \cite{2014ApJ...789L..19F}; \cite{2015ApJ...798L..10L}; \cite{2018ApJ...852...98W}) and their results support this theory.

Approximately ten years after the Yohkoh satellite discovery, small anemone jets were found in the chromosphere by observation with the Hinode satellite (\cite{2007Sci...318.1591S,2008ApJ...683L..83N,2011PhPl...18k1210S}).
Although the spatial scale and velocity of the anemone jets differ significantly between the corona and chromosphere (see table \ref{anemone_ta}), their observational features and numerical simulations  (\cite{2008ApJ...683L..83N,2013PASJ...65...62T}) support the fact that chromospheric anemone jets are also explained by magnetic reconnection. 
In both coronal and chromospheric cases, the jets speed is approximately the Alfv\'en speed, and  the lifetime normalized by the Alfv\'en time is approximately $10-100$. 
These properties are consistent with reconnection theory and suggest that anemone jets are induced by reconnection regardless of their spatial scale.
This fact reflects the scale universality of magnetohydrodynamics (MHD).

From the MHD scale universality, it can be expected that anemone jets having physical values as indicted in table \ref{anemone_ta} occur in the solar photosphere.
However, considering that the solar photosphere is not magnetically dominant (plasma beta $\beta\sim1-10$),  
it is not apparent that jets can be created as a result of magnetic reconnection, as in the chromosphere and corona of $ \beta \ll1 $.
\citet{2016A&A...596A..36P} performed 3D MHD simulation in an environment of $\beta\simeq 1$; however, this calculation did not include gravity, and hence it could possibly not apply to the dynamics in the photosphere.
Moreover, owing to the lack of spatial resolution, observing jets with a size of approximately $100\,\mathrm{km}$ remains a challenge.
Several cases of brightening with inverted-Y-shaped loops in the upper photosphere and the lower chromosphere have been reported (\cite{2011ApJ...736L..35Y,2017A&A...597A.127B,2017ApJS..229....4C,2018ApJ...854...92T,2019ApJ...883..115N}); however, their length is approximately $1000\,\mathrm{km}$.
The jet phenomenon of approximately $100\,\mathrm{km}$ has not yet been observed.

The presence of photospheric anemone jets could be valuable as a source of waves in the lower solar atmosphere.
When a photospheric anemone jet is generated, the surrounding pressure and magnetic field fluctuate and MHD waves are generated.
Because the solar atmosphere is stratified by gravity, these MHD waves propagate to the upper layer with growing amplitude and form shock waves, which could drive spicules and surges.
Such a growth process of MHD waves is the same mechanism that drives spicules by p-mode leakage and slow mode MHD waves generated from nonlinear Alfv\'en waves (\cite{1982SoPh...75...99S,2004Natur.430..536D,1982SoPh...75...35H,1999ApJ...514..493K,2020ApJ...891..110W}) or  surges from Ellerman bombs (e.g., \cite{2014ApJ...790L...4Y}).
As a model to explain the spicule formation with the same mechanism as coronal jets, \citet{2016ApJ...828L...9S} proposed that spicules are formed by the ejection of microfilament.
However, if the microfilament ejected at Alfv\'en speed is not re-accelerated, the height that can be achieved is $\sim H/\beta$ ($H$: scale height$\sim150\,\mathrm{km}$ in the solar photosphere).
Therefore, it is difficult for the microfilament itself to achieve the height of the spicule even if the microfilament erupts from the photosphere or lower chromosphere with $\beta\sim1$.
Hence, even when considering microfilament ejection, it is necessary to consider the MHD waves generated by the ejection or the re-acceleration mechanism of the microfilament. 
Moreover, magnetic reconnection has been proposed as the origin of the waves necessary to cause chromospheric and coronal heating (e.g., \cite{1991ApJ...372..719P}).
If reconnection events in the lower atmosphere provide waves to heat, photospheric jets could be a candidate.
For these reasons, we must study photospheric anemone jets to determine spicule models and clarify the heating mechanism of the solar atmosphere.

In this study, we perform a 3D MHD simulation of magnetic reconnection with solar photospheric parameters to investigate the properties of photospheric anemone jets.
Then, we discuss the degree to which MHD waves are generated from the jets and how they influence the upper atmosphere.

\begin{table*}[]
   \tbl{%
           Typical physical quantities of anemone jets in each layer of the solar atmosphere.
           }{%
   \begin{tabular}{c c c c c c} \hline
     Region & Length $(\mathrm{km})$ &Velocity $(\mathrm{km}\,\mathrm{s}^{-1})$ &Lifetime $(\mathrm{s})$& Alfv\'en velocity  $(\mathrm{km}\,\mathrm{s}^{-1})$ & $t/t_A$  \\ \hline
     Corona & $10^4-10^5$ & $100-10^3$ & $10^3-10^4$ & $100-10^3$ &$10-100$\\
     Chromosphere & $10^3-10^4$ & $10-100$ & $100-10^3$  & $10-100$ & $10-50$\\
     Photosphere? & $10-100$ & $1-10$ & $10-100$& $1-10$ &$10?$\\ \hline
   \end{tabular}}
  \label{anemone_ta}
\end{table*}

\section{Method}

Although in the solar photosphere the temperature is $\sim 6000\,\mathrm{K}$, ionization degree is low, and majority of components of the fluid are neutral, numerous collisions do occur between the neutral and plasma owing to the high density in the photosphere. 
The collision time between the ion and neutral $\tau$ represents $\nu_{in}^{-1}\simeq n_n\surd(8k_B T/\pi m_{in})\Sigma_{in}$, where $\nu_{in}$, $n_n$, $m_{in}=(m_i+m_n)/m_im_n$, and  $\Sigma_{in}\simeq5\times 10^{-15}\,\mathrm{cm}^2$ are the collisional frequency of ions with the neutrals, number density of the neutrals, reduced mass of the ion and neutral, and ion-neutral collision cross-sections (\cite{2006A&A...450..805L}). 
We assume that all the elements are hydrogen and $n_n = 10^{17}\,\mathrm{cm}^{-3}$ in the photosphere; then, $m_{in}$ is the proton mass $m_p$ and $\tau\simeq 10^{-9}\,\mathrm{s}$.
This time is considerably shorter than the time resolution of our simulation $dt\sim 0.1\times0.1\,\mathrm{km}/(c_s= 8.15\,\mathrm{km}\,\mathrm{s}^{-1})\sim 10^{-3}\,\mathrm{s}$.
Thus, we can neglect the effect of the partially ionized plasma in our simulation, and we adopt the one fluid (MHD) approximation.

     For our numerical simulation, we use Athena++ code with the van Leer predictor-corrector scheme and Piecewise Linear Method \citep{2020arXiv200506651S}.
We solve the ideal MHD equations including uniform gravity and simple form radiative cooling (Newton cooling).
The basic equations are as follows.
\begin{eqnarray}
\lefteqn{\frac{\partial \rho}{\partial t} + \bm{\nabla}\cdot(\rho \bm{v})= 0 }      \label{eq:mass}      \\
&&\frac{\partial \rho \bm{v}}{\partial t} + \bm{\nabla}\cdot(\rho\bm{v}\bm{v}) = -\bm{\nabla}p + \frac{1}{4\pi}(\bm{\nabla}\times\bm{B})\times\bm{B}+\rho \bm{g}      \label{eq: eom}   \\
&&\frac{\partial (e + \frac{1}{2}\rho v^2 + \frac{B^2}{8\pi})}{\partial t} + \bm{\nabla}\cdot \bigl[(h + \frac{1}{2}\rho v^2 + \frac{B^2}{4\pi})\bm{v} 
- \frac{1}{4\pi}(\bm{B}\cdot\bm{v})\bm{B} )\bigr] \nonumber\\
&&=\rho \bm{g} \cdot\bm{v} -\frac{\rho R}{\mu (\gamma -1)}\frac{T-T_0}{\tau_{\mathrm{cooling}}}     \label{eq: energy}    \\
&&\frac{\partial \bm{B}}{\partial t} =\bm{\nabla}\times(\bm{v}\times\bm{B})  \label{eq: ind}   \\
&&p = \frac{\rho R T}{\mu} \label{eq: eos}
\end{eqnarray} 
, where $e$ is the internal energy of the fluid, $h=p+e$ is the enthalpy, $T_0=6000\,\mathrm{K}$ is uniform initial temperature, $\tau_{\mathrm{cooling}}=1\,\mathrm{s}$ is mean cooling time and $\mu=1.25$ is mean molecular weight in the solar photosphere.   
Note that $\tau_{\mathrm{cooling}}=1\,\mathrm{s}$ agrees with the cooling time deduced for the photosphere with temperature $T=6000\,\mathrm{K}$ (\cite{1976ASSL...53.....A}, p. 474 table X-4), and that it decreases with height in the actual sun.
In our simulation, because magnetic reconnection occurs at a height of approximately $20\,\mathrm{km}$ and the majority of the temperature change occurs in this vicinity, we set the cooling time uniform within the simulation box for simplicity.
$\bm{g}$ is the gravitational acceleration of the solar atmosphere ($|\bm{g}| = 2.7\times10^4\,\mathrm{cm}\,\mathrm{s}^{-2}$) and we set $\bm{g}=(g_x,g_y,g_z) =(-g,0,0)$.
We assume a specific heat ratio of $\gamma=5/3$.

For the normalization of the simulation, we use the typical length $L_0=10\,\mathrm{km}$, initial temperature $T_{0}= 6000\,\mathrm{K}$, and density in the bottom boundary $\rho_0 = 1.0\times10^{-7}\,\mathrm{g}\,\mathrm{cm}^{-3}$.
Then, we use the unit of velocity, pressure, magnetic field, and time as $v_0=\surd(RT_0/\mu)=6.32\times10^5\,\mathrm{cm}\,\mathrm{s}^{-1}, p_0=\rho_0v_{0}^2=3.97\times10^4\,\mathrm{erg}\,\mathrm{cm}^{-3}, B_0=\surd(4\pi p_0)=706\,\mathrm{G}$, and $t_0=L_0/v_{0}=1.57\,\mathrm{s}$ (see table \ref{unit_ta}).
Note that $v_0$ is different from the sound speed in the solar photosphere $c_{s0}=\surd(\gamma RT_0/\mu)=8.15\times10^5\,\mathrm{cm}\,\mathrm{s}^{-1}$.

\begin{table*}[]
\tbl{Units used in numerical simulation. 
\label{bipole}}{%
\begin{tabular*}{140mm}{@{\extracolsep{\fill}}c c c c c c c} 
\hline  
   Length& Density& Temperature &Velocity&Gas pressure & Magnetic field&Time  
 \\ \hline
     $L_0$ & $\rho_0$ & $T_0$  & $v_0$ & $p_0$ &$B_0$ & $t_0$\\
     $10\,\mathrm{km}$ &  $10^{-7}\,\mathrm{g}\,\mathrm{cm}^{-3}$ & $6000\,\mathrm{K}$  & $6.32\times10^{5}\,\mathrm{cm}\,\mathrm{s}^{-1}$ &  $3.97\times10^{4}\,\mathrm{erg}\,\mathrm{cm}^{-3}$& $706\,\mathrm{G}$ & $1.57\,\mathrm{s}$ \\ [3pt]
  \hline\noalign{\vskip3pt} 
\end{tabular*}}
 \label{unit_ta}
\end{table*}

As an initial condition, we assume hydrostatic equilibrium with uniform temperature $T_0$.
Moreover, we present a potential field as follows (\cite{2018ApJ...852...98W}).
\begin{eqnarray}
\lefteqn{\bm{B} =   (c_1\cos\theta,c_1\sin\theta,0)  +   \sum_{i=1,16} \bm{\nabla}\times \bm{A}_i}        \label{eq:ini_B}       \\
  &&\bm{A}_i = \frac{b_i x_i^3}{2[ x_i^{\prime 2} + (y_i^{\prime } - y_c)^2 + z_i^{\prime 2}]^{3/2}}
   \times[-z_i^{\prime}\bm{e}_y +(y_i^{\prime} - y_c)\bm{e}_z]              \label{eq:potential}      
\end{eqnarray} 
$c_1= -0.8$ and $\theta=-22^{o}$ are the parameters of the background field strength and angle, respectively.
$x_i^{\prime}=x - x_i, y_i^{\prime}=y - y_i, z_i^{\prime}=z - z_i$.
We present the values  of $b_i, x_i, y_i, z_i$ in table\ref{bipole}.  
Figure 1 (a) displays this initial field.

\begin{figure*}[]
  \begin{center}
  \includegraphics[bb= 0 0 585 236,width=160mm]{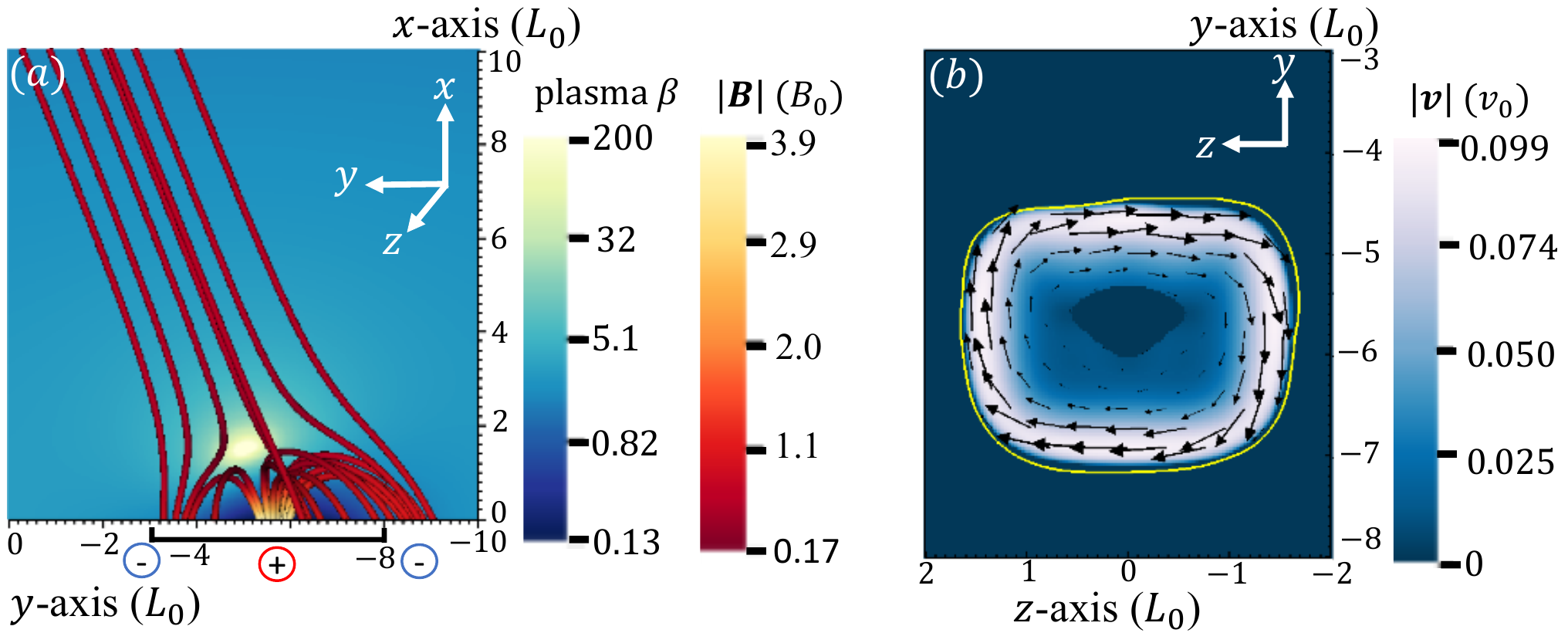}
  \end{center}
  \caption{%
     (a): Initial potential field and distribution of plasma $\beta$. The symbols in the bottom boundary mean the polarity of the magnetic field. 
     The region indicated by the black line along the y-axis represents the range of the (b).
     (b): Velocity distribution in the bottom boundary at $t=60t_0$. Yellow contour means polarity inversion line.
  }%
  \label{fig:initial_B}
\end{figure*}

\begin{table}
\tbl{Parameters of magnetic field. 
\label{bipole}}{%
\begin{tabular*}{70mm}{@{\extracolsep{\fill}}c c c c c} 
\hline  
   $i$ & $b_i$ & $x_i$ &$y_i$ &$z_i$  
 \\ \hline
     1 & -1.35 & -1.0  & -0.5 & -1.0\\
     2 & -1.35 & -1.0  & -0.5 & -0.5\\
     3 & -1.35 & -1.0  & -0.5 & 0\\
     4 & -1.35 & -1.0  & -0.5 & 0.5\\
     5 & -1.35 & -1.0  & -0.5 & 1.0\\
     6 & -1.35 & -1.0  & 0 & -1.0\\
     7 & -1.35 & -1.0  & 0 & 0\\
     8 & -1.35 & -1.0  & 0 & 1.0\\
     9 & -1.17 & -1.0 & -1.5 & -1.0\\ 
     10 & -1.17 & -1.0 & -1.5 & -0.5\\ 
     11 & -1.17 & -1.0 & -1.5 & 0\\ 
     12 & -1.17 & -1.0 & -1.5 & 0.5\\ 
     13 & -1.17 & -1.0 & -1.5 & 1.0\\ 
     14 & -1.17 & -1.0 & -1.0 & -1.0\\ 
     15 & -1.17 & -1.0 & -1.0 & 0\\ 
     16 & -1.17 & -1.0 & -1.0 & 1.0  \\ [2pt]
  \hline\noalign{\vskip3pt} 
\end{tabular*}}
\begin{tabnote}
\hangindent6pt\noindent
\hbox to6pt{\footnotemark[$*$]\hss}\unskip%
Meaning of $b_i$,  $x_i$,  $y_i$, and $z_i$ is provided in the text.  
\end{tabnote}
\end{table}

We set the velocity field in the bottom boundary as follows (\cite{2009ApJ...691...61P,2018ApJ...852...98W}):
\begin{eqnarray}
\lefteqn{v_x = 0 }      \label{eq:v_x}       \\
\bm{v}_{\perp}& =&  f(t)\nu_0 g(B_x)\bm{e}_x\times\bm{\nabla}B_x           \label{eq:vnorm} \\
g(B_x) &=& k\frac{B_r - B_l}{B_x}\tanh\left(k\frac{B_x-B_l}{B_r-B_l}\right)\quad B_l\leq B_x\leq B_r\\
      \quad    \quad            &=& 0      \quad   \verb|otherwise|     \label{eq:ggg}.
\end{eqnarray}  
We set $k=4.0$, $\nu = 0.002$, $B_l = 0.15$, and $B_r = 3.6$.   
$f(t)$ is the time development of the bottom boundary photospheric motion, and we assume the following:
\begin{eqnarray}
f(t)& =& t/t_1   \quad (t<t_1)        \label{eq:t1} \\
 &=& 1 \quad (t_1\le t<t_2)        \label{eq:tt}\\
  &=& 1 - \frac{t-t_2}{t_1}      \quad  (t\le t_2)     \label{eq:t2}   
\end{eqnarray}  
, where $t_1=60t_0$ and $t_2=75t_0$.
Note that although the adopted function is the same, the values of these parameters are different from those of the previous study, resulting in different values of the velocity field from those of the previous study.
These values are based on the assumption that granulation satisfies the Kolmogorov law $\epsilon\sim v_{\lambda}^3/\lambda$.
Specifically, the typical size of the granulation is $1000 \,\mathrm{km}$ and the speed is $ 1 \, \mathrm{km}\,\mathrm{s}^{-1} $; then, the speed of the smaller vortex of approximately $10\,\mathrm{km}$ is determined from the Kolgomogorov law as follows:
\begin{equation}
v_{\lambda=10\,\mathrm{km}}\sim v_{\lambda=1000\,\mathrm{km}} \left(\frac{\lambda=10\,\mathrm{km}}{\lambda=1000\,\mathrm{km}}\right)^{1/3}\sim 0.25\,\mathrm{km}\,\mathrm{s}^{-1} .\label{eq:kolgo_v}
\end{equation}
Moreover, the time scale of the $10\,\mathrm{km}$ scale small vortex is  
\begin{equation}
t_{\lambda=10\,\mathrm{km}}\sim \frac{\lambda=10\,\mathrm{km}}{v_{\lambda=10\,\mathrm{km}}}\sim 40\,\mathrm{s}. \label{eq:kolgo_t}
\end{equation}
These values are consistent with the velocity and lifetime used in our simulation within a factor three.
Figure 1 (b) displays this boundary motion.
For gas pressure and density, the initial values are maintained, and for the magnetic field, the gradient is set to zero.
In the other boundary, we use an open boundary and set the gradient of the physical value as zero.

We set our simulation box as $[x_{\mathrm{min}},x_{\mathrm{max}}]\times[y_{\mathrm{min}},y_{\mathrm{max}}]\times[z_{\mathrm{min}},z_{\mathrm{max}}]=[0,50L_0]\times[-12.5L_0,7.5L_0]\times[-5L_0,5L_0]$.
We set the grid size as $dx=dy=dz=0.05$.
Moreover, the lattice spacing is halved at $[0,10L_0]\times [-12.5L_0, -2.5L_0] \times [-2.5L_0,2.5L_0]$ to sufficiently resolve the anemone-type magnetic field.

\section{Results}

\begin{figure*}[]
  \begin{center}
  \includegraphics[bb= 0 0 412 369,width=160mm]{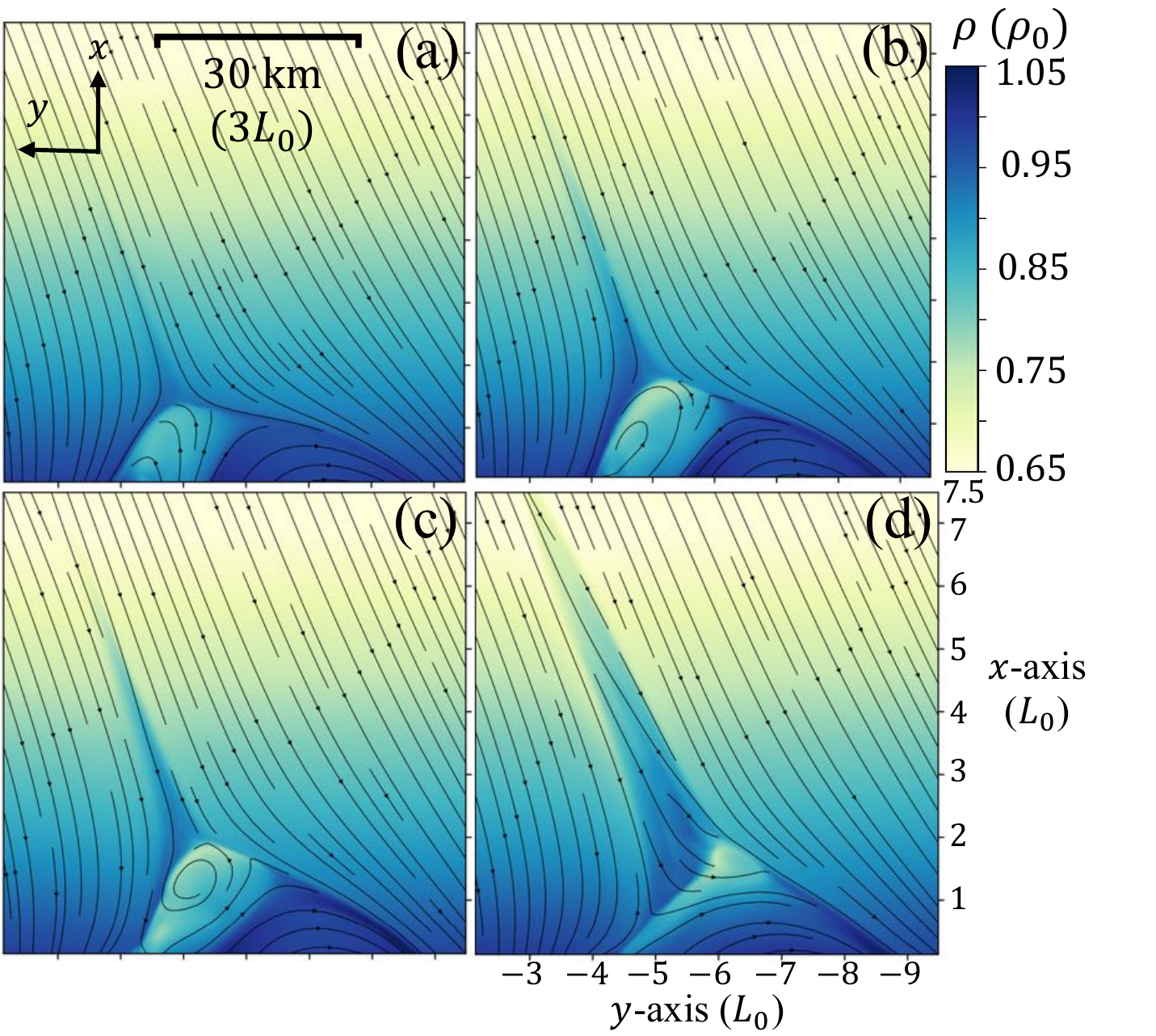}
  \end{center}
  \caption{%
    Time development of density distribution in $z=0$ plane.
    (a): $t=60t_0$, (b): $t=70t_0$, (c): $t=80t_0$, (d): $t=100t_0$.
    Black lines display magnetic field lines integrated by $x, y$ components in the plane.
    The upper boundary is the $x=7.5L_0$ plane, where we measure the MHD wave energy fluxes in figures \ref{fig:energy_flux} and \ref{fig:flux_plot}.
    (An animation of this figure is available.)
 }%
  \label{fig:2d_rho}
\end{figure*}

  Figure \ref{fig:2d_rho} displays the time evolution of the density distribution on the $z=0$ plane and magnetic field lines integrated in the plane.
First, the initial magnetic fields are twisted to form a sheared loop because of the photospheric motion based on the boundary condition (figure \ref{fig:2d_rho}a).  
As the magnetic pressure increases owing to the magnetic energy injected by the photospheric motion, the sheared loop rises (figure \ref{fig:2d_rho}b). 
At this time, magnetic reconnection occurs at the top of the sheared loop, and a jet-like structure, which is an elongated density increase along the magnetic field lines and appears as a jet in imaging observations, is generated.
This reconnection removes the magnetic field that holds the sheared loop from above and thus further promotes the rising of the sheared loop.
When the twisted sheared loop emerges sufficiently, the twisted sheared loop and the background field cause magnetic reconnection, resulting in an untwisting jet-like structure, which releases numerous Alfv\'en waves (figures \ref{fig:2d_rho}c and  \ref{fig:2d_rho}d).
This mechanism is similar to coronal jet simulation (e.g., \cite{2009ApJ...691...61P}; \cite{2013ApJ...769L..21A}; \cite{2018ApJ...852...98W}), which assumes a corona of $\beta\ll1$.

The length, apparent speed, and lifetime of this jet-like structure are approximately $100\,\mathrm{km}$, $5\,\mathrm{km}\,\mathrm{s}^{-1}$, and $80\,\mathrm{s}$.
This apparent speed approximately corresponds to Alfv\'en speed at this point.
The length of the twisted sheared loop is approximately $30\,\mathrm{km}$, and hence the Alfv\'en time $t_A$ is $t_A \sim 30\,\mathrm{km}/5\,\mathrm{km}\,\mathrm{s}^{-1}\sim 6\,\mathrm{s}$.
Thus, the lifetime $t$ divided by the Alfv\'en time $t_A$ is $t/t_A\sim 80\,\mathrm{s}/6\,\mathrm{s}\sim 13.3$.
All of these values are consistent with those in table \ref{anemone_ta}.

\begin{figure*}[]
  \begin{center}
  \includegraphics[bb= 0 0 720 625,width=160mm]{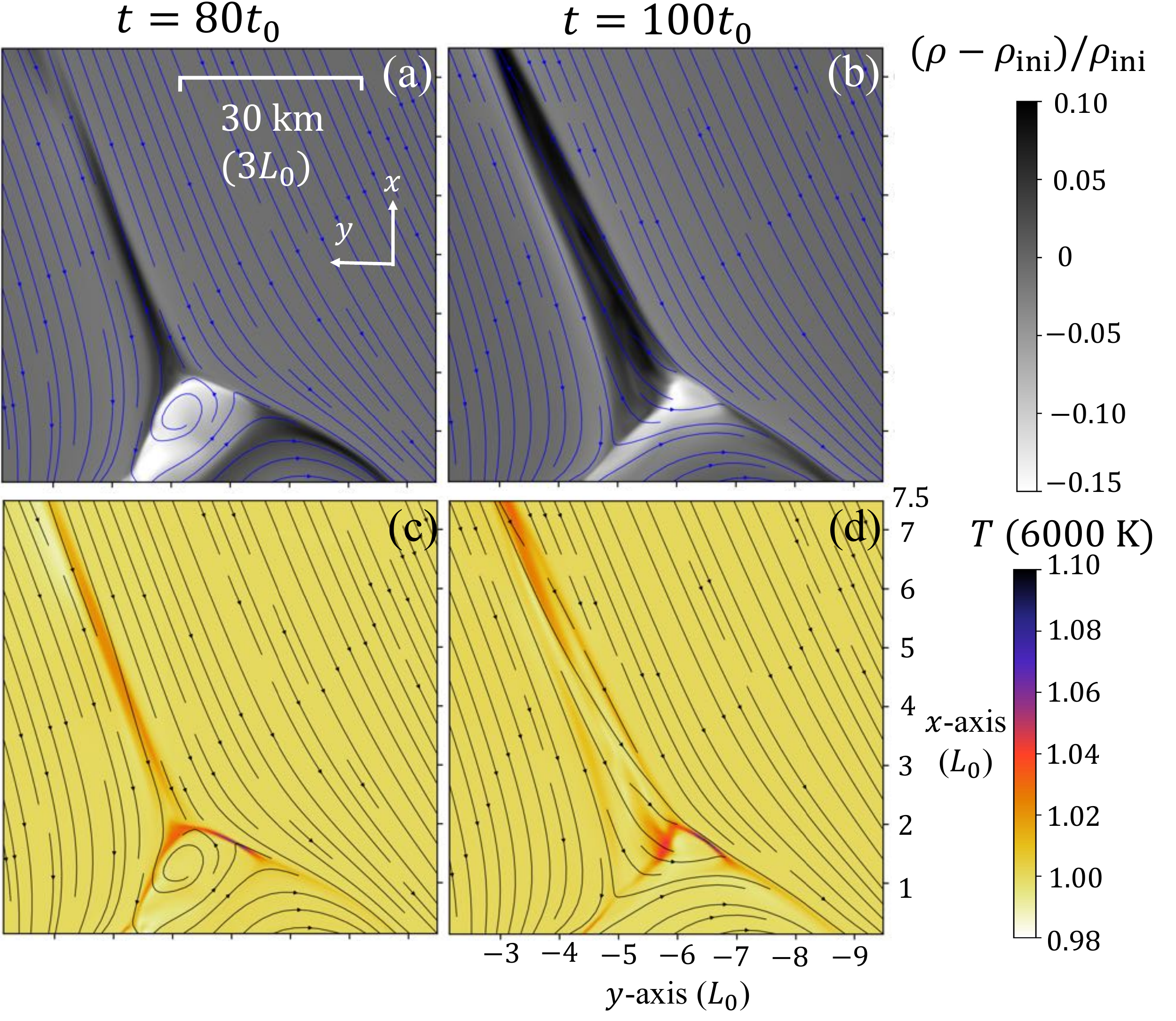}
  \end{center}
  \caption{%
Rate of increase in density and temperature of the jet-like structure  in $z=0$ plane.
The left and right columns indicate the same time as in figures \ref{fig:2d_rho}c and \ref{fig:2d_rho}d, respectively.
$\rho_{\mathrm{ini}}$ is the initial density.
The upper boundary is the $x=7.5L_0$ plane, where we measure the MHD wave energy fluxes in figures \ref{fig:energy_flux} and \ref{fig:flux_plot}.
 }%
  \label{fig:2d_jet_like}
\end{figure*}

Figure \ref{fig:2d_jet_like} indicates how much the density and temperature increase in the jet-like structure.
Figures \ref{fig:2d_jet_like}a and \ref{fig:2d_jet_like}b indicate that the jet-like structure has a density of approximately 1.1 times greater than its surroundings.
Furthermore, figures \ref{fig:2d_jet_like}c and \ref{fig:2d_jet_like}d indicate that the temperature increases approximately 200 K in the jet-like structure, and that the temperature increase has an inverted-Y shape.

\begin{figure*}[]
  \begin{center}
  \includegraphics[bb= 0 0 811 814,width=160mm]{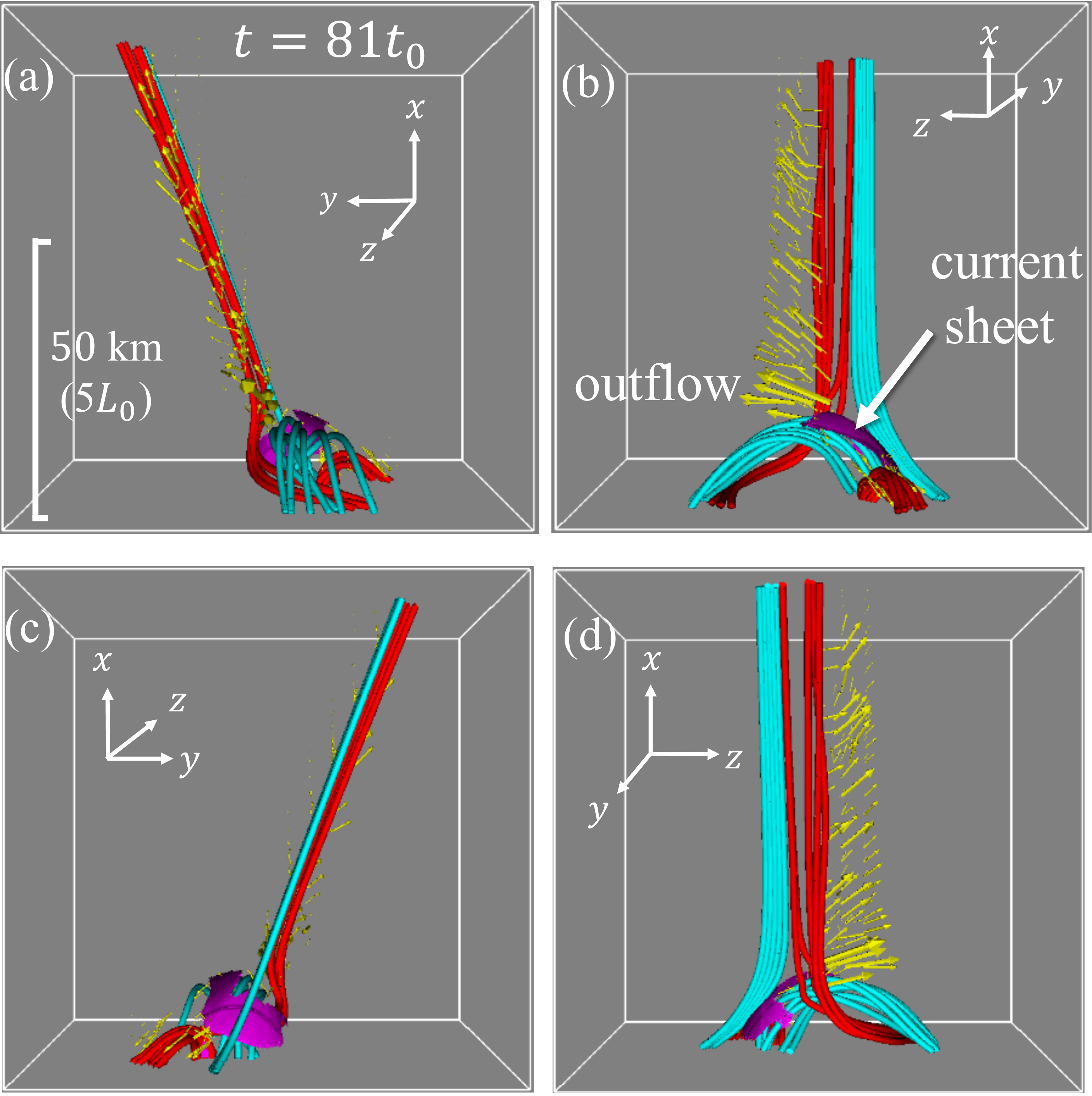}
  \end{center}
  \caption{%
  3D view of the jet-like structure in $t=81t_0$.
  Yellow arrows and pink surface are the velocity and current sheet.
  Light blue lines indicate twisted sheared loop and open fields before magnetic reconnection.
  Red lines indicate reconnected open fields and post-flare loop.
  The size of the largest arrow corresponds to $0.6v_0$.
 }%
  \label{fig:3d_jet}
\end{figure*}

     Figure \ref {fig:3d_jet} displays the 3D appearance at $t=81t_0$.
From figure \ref{fig:3d_jet}b, we can observe that magnetic reconnection occurs between the twisted sheared loop and background fields, and plasma frozen in the reconnected field lines is accelerated approximately in the z-direction; that is, in the direction normal to the ambient field.
Moreover, the twisted loop, background field, and current sheet are aligned in the z-direction.
Therefore, the reconnection process can be investigated in more detail with an appropriate “$y=\mathrm{const}$” cross section, though it is actually a complex 3D phenomenon.

\begin{figure*}[]
  \begin{center}
  \includegraphics[bb= 0 0 964 1239,width=160mm]{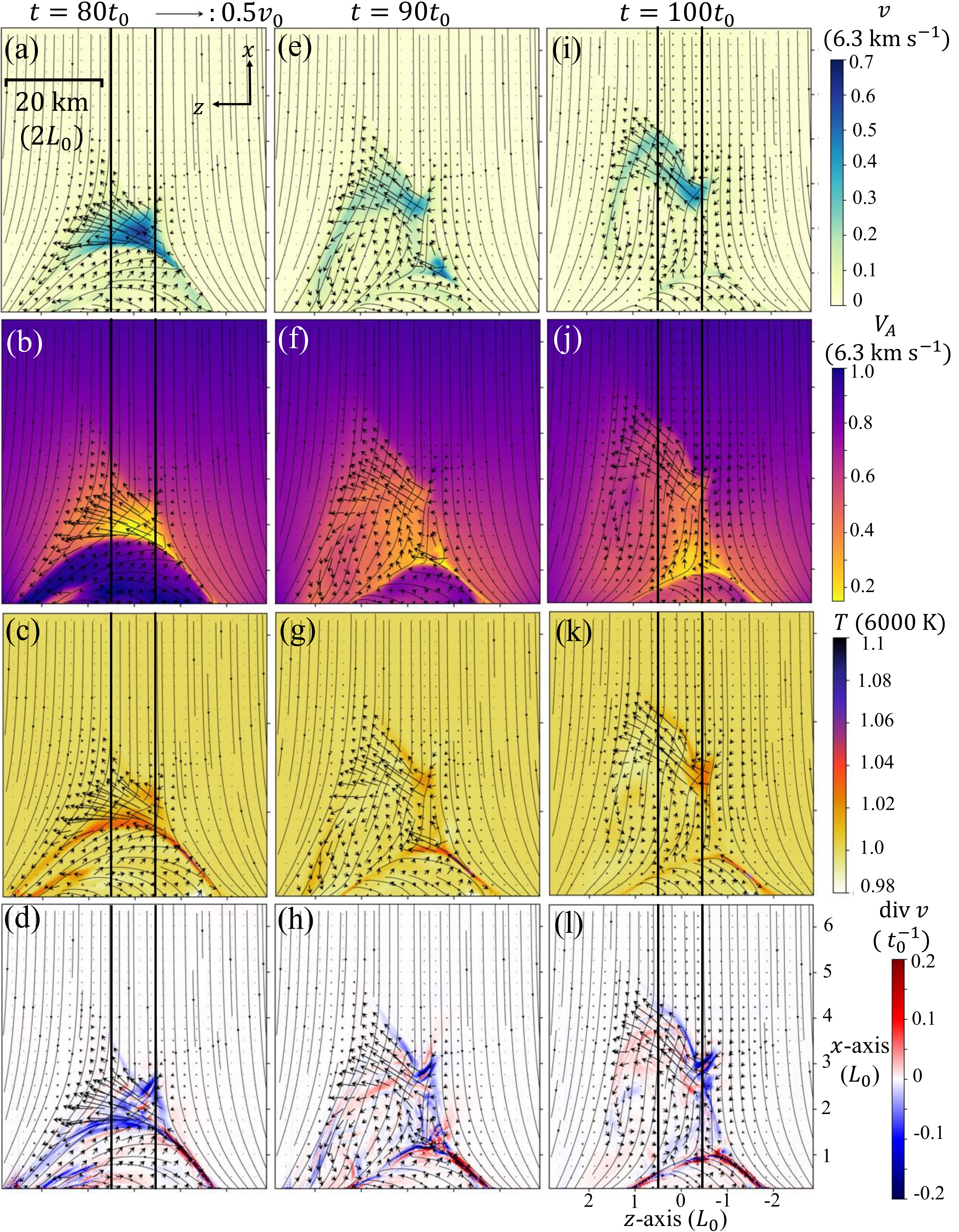}
  \end{center}
  \caption{%
    Physical values in $y=-4.985L_0$ plane at $t=80t_0$ ((a)--(d)), $t=90t_0$ ((e)--(h)), and  $t=100t_0$ ((i)--(l)).
   Each row shows velocity ((a), (e), and (i)), Alfv\'en velocity  ((b), (f), and (j)), temperature  ((c), (g), and (k)) and divergence of velocity ((d), (h), and (l)).
   Black arrows are velocity  in the plane.
    Black lines show magnetic field lines integrated by $x, z$ components in the plane.
 One dimensional distributions of some physical quantities along two black solid lines in $t=80t_0$ and $100t_0$ are shown in figure \ref{fig:revise_1d}.
 (An animation of (d), (h), and (l) is available.)
 }%
  \label{fig:reco_rate}
\end{figure*}
 
 \begin{figure*}[]
  \begin{center}
  \includegraphics[bb= 0 0 964 958,width=160mm]{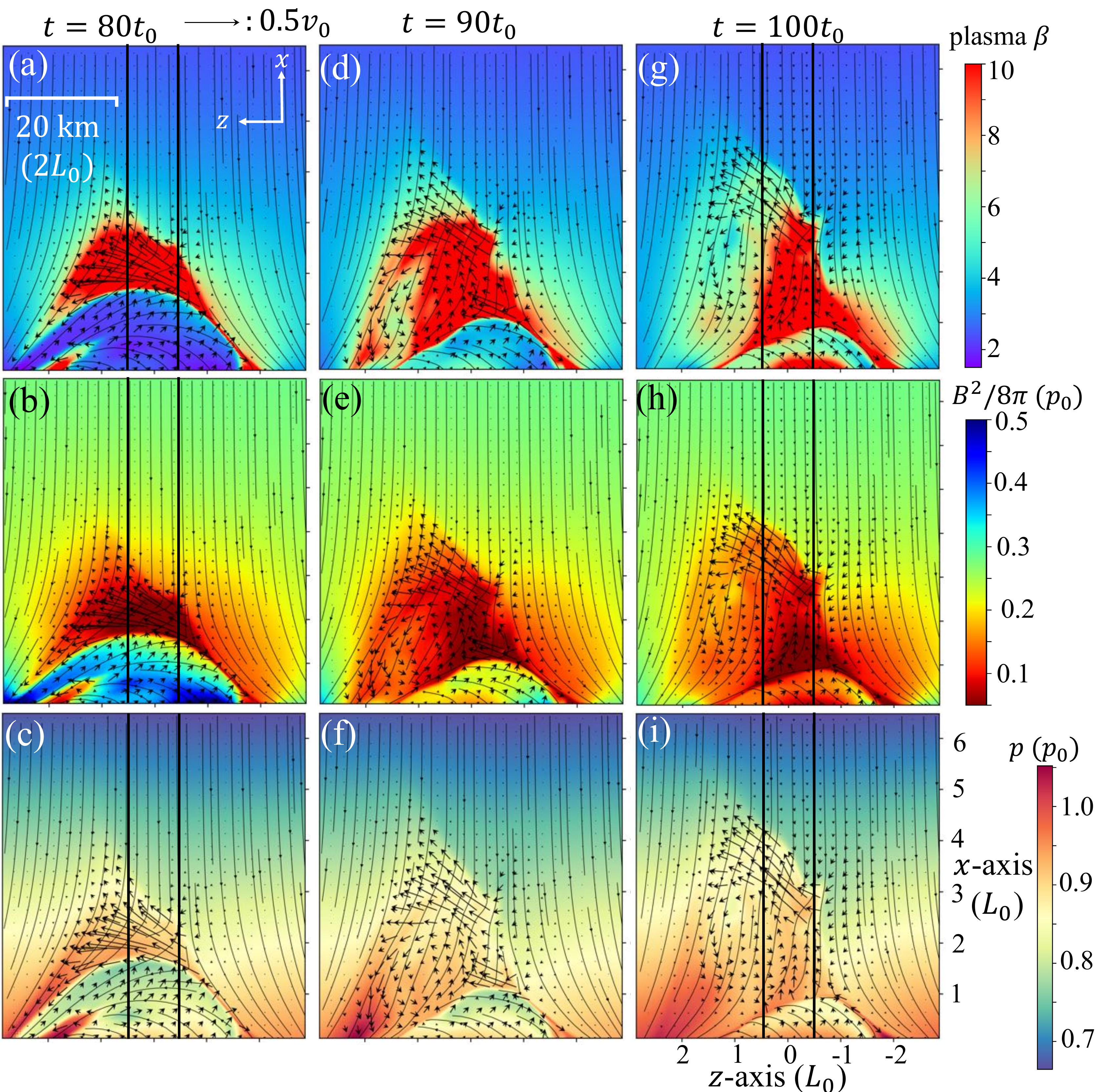}
  \end{center}
  \caption{%
    Physical values in $y=-4.985L_0$ plane at $t=80t_0$ ((a)--(c)), $t=90t_0$ ((d)--(f)), and $t=100t_0$ ((g)--(i)).
   Each row shows plasma $\beta$ ((a), (d), and (g)), magnetic pressure ((b), (e), and (h)), and gas pressure  ((c), (f), and (i)).
   Black arrows are velocity  in the plane.
    Black lines show magnetic field lines integrated by $x, z$ components in the plane.
    One dimensional distributions of some physical quantities along two black solid lines in $t=80t_0$ and $100t_0$ are shown in figure \ref{fig:revise_1d}.
 }%
  \label{fig:reco_rate_2}
\end{figure*}

 \begin{figure*}[]
  \begin{center}
  \includegraphics[bb= 0 0 947 682,width=160mm]{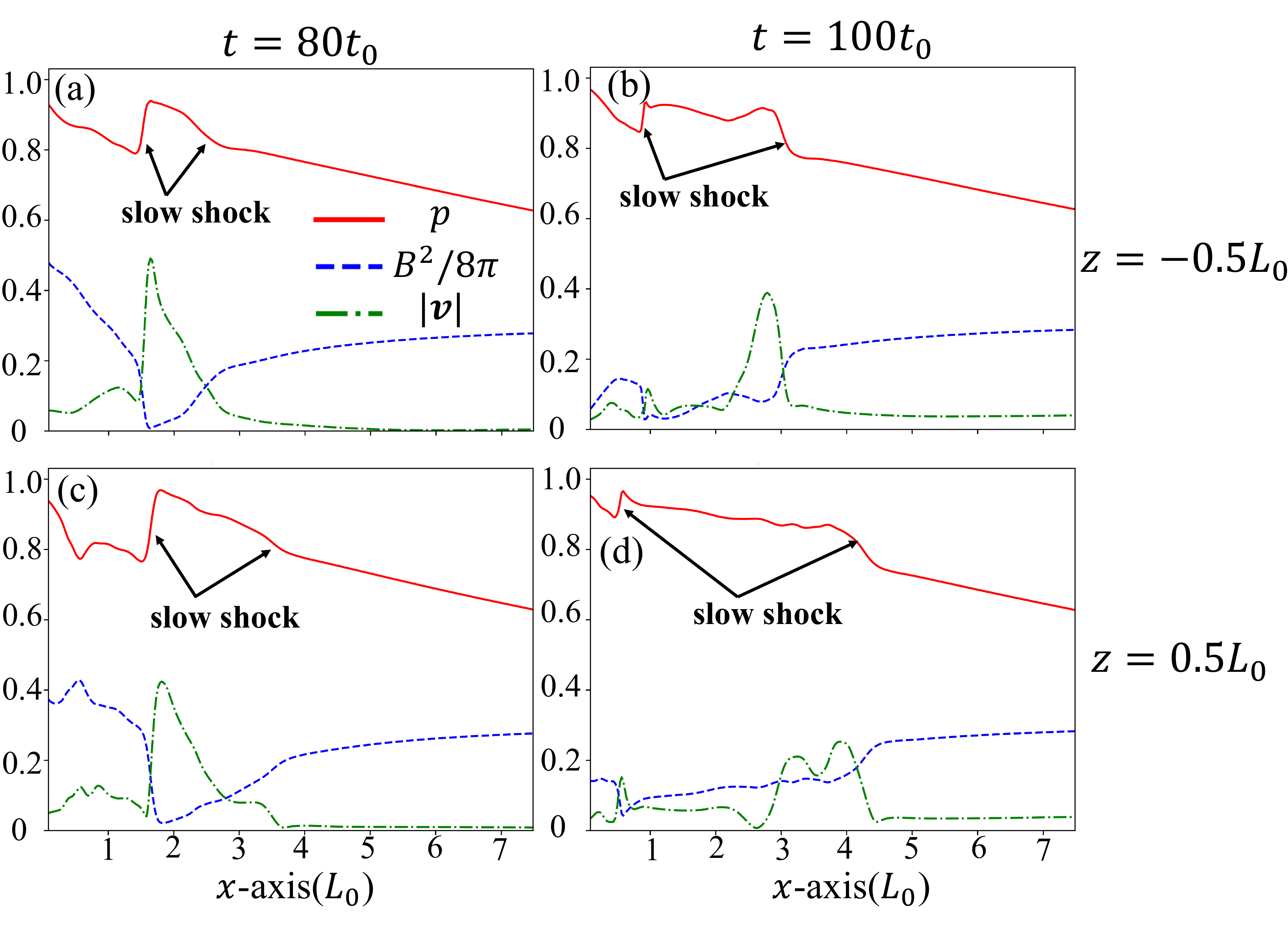}
  \end{center}
  \caption{%
        One-dimensional (1D) distribution of gas pressure, magnetic pressure, and velocity along the solid black lines at $z=0.5$ and $-0.5$ shown in figures \ref{fig:reco_rate} and \ref{fig:reco_rate_2}.
The solid red lines indicate gas pressure, the dashed blue lines indicate magnetic pressure, and the dash-dot green lines indicate speed.
The left column shows the distributions at $t=80$, and the right column shows the distributions at $t=100$.
(a) and (b) correspond to distributions at z = -0.5, and (c) and (d) correspond to distributions at $z = 0.5$.
The units are $3.97\times10^4\,\mathrm{erg}\,\mathrm{cm}^{-3}$ for gas and magnetic pressure and $6.32\times10^5\,\mathrm{cm}\,\mathrm{s}^{-1}$ for plasma speed.
}%
  \label{fig:revise_1d}
\end{figure*}

Figures \ref{fig:reco_rate} and \ref{fig:reco_rate_2} display the physical values of the jet-like structure in the $y= -4.95L_0$ plane, where we can interpret the process of magnetic reconnection in a broadly 2D view.
Closed fields at the bottom indicate the twisted sheared loop.
From figure \ref{fig:reco_rate}a, we can observe reconnection inflow and outflow.
The outflow's velocity is $\sim 0.7v_{0}$, which corresponds to Alfv\'en velocity  in the sheared loop (figure \ref{fig:reco_rate}b).
The inflow's velocity is $\sim 0.1v_{0}$.
Then, we can estimate the reconnection rate $v_{\mathrm{in}}/v_{\mathrm{out}}\sim0.1$.
From figure \ref{fig:reco_rate}c, we can observe that the maximum temperature in the current sheet is $\sim6600 \mathrm{K}$ and the outflow temperature is $\sim6300 \mathrm{K}$.
From figure \ref{fig:reco_rate}d, we can observe the area of $\mathrm{div}v<0$ between the inflow and outflow, which corresponds to the compression occurring.
Moreover, the gas pressure becomes stronger and the magnetic field becomes weaker before and after the $\mathrm{div}v<0$ region (figures \ref{fig:reco_rate_2}b and \ref{fig:reco_rate_2}c), which indicates that the $\mathrm{div}v<0$ corresponds to a slow shock formed by magnetic reconnection.
The slow shock and the reconnection rate in our simulation imply that this reconnection has similar properties to those of the Petschek model \citep{1964NASSP..50..425P}.

\begin{figure*}[]
  \begin{center}
  \includegraphics[bb= 0 0 1347 1419,width=170mm]{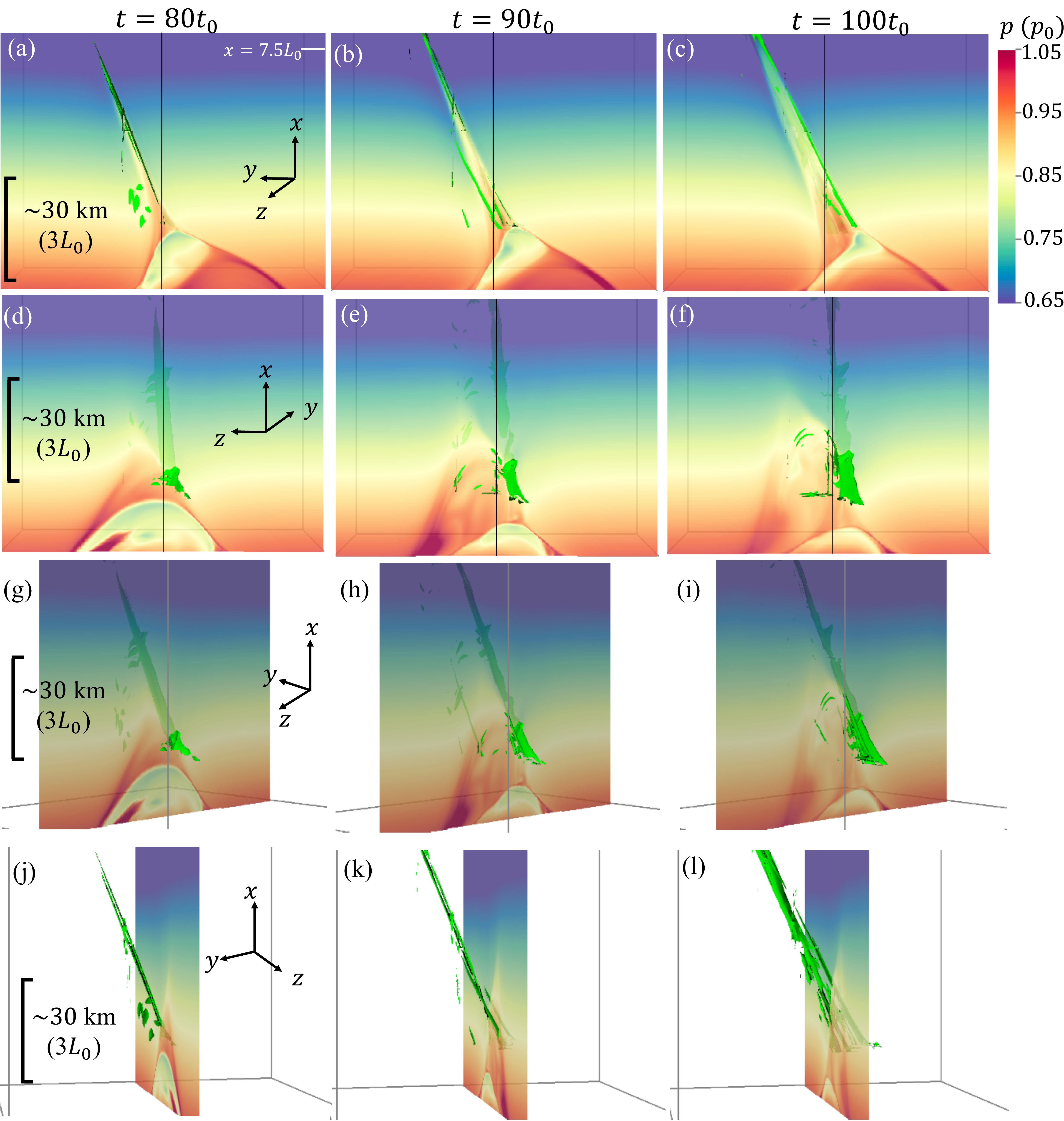}
  \end{center}
  \caption{%
 Time development of the $\mathrm{div}v<0$ region and the gas pressure distribution shown in 2D planes.
 In each figure, the green surface indicates the regions of $\mathrm{div}v=-0.09t_0^{-1}$ and $-0.07t_0^{-1}$.
 (a)--(c): Gas pressure in the $z=0$ plane.
 The solid black line shown in the center of each figure shows the $y=-4.95L_0$ plane.
  (d)--(f): Gas pressure in the $y=-4.95L_0$ plane.
 The solid black line shown in the center of each figure shows the $z=0L_0$ plane.
  (g)--(i): Gas pressure in the $y=-4.95L_0$ plane and the 3D isosurfaces of $\mathrm{div}v$ from an oblique view.
  }%
  \label{fig:slow_current}
\end{figure*}

From figures \ref{fig:reco_rate}e--h, we can observe that the slow shock is displaced upward and the plasma is accelerated approximately in the z-direction in the upper atmosphere.
This can certainly be confirmed by the one-dimensional (1D) distribution displayed in figure \ref{fig:revise_1d}, which indicates both the upward and downward slow shocks formed by reconnection.
Note that this displacement of the slow shock is not a result of the upward propagation of the slow shock in this plane.
Figure \ref{fig:slow_current} presents the time development of the $\mathrm{div}v<0$ region and gas pressure in the $z=0$ and $y=-4.95L_0$ planes.
From this figure, we can observe that the region of $\mathrm{div}v<0$ exists along the direction of the jet-like structure, that is, a uniform background magnetic field. 
Moreover, the strong $\mathrm{div}v<0$ region extends from near the region where magnetic reconnection is occurring.
These facts indicate that the $\mathrm{div}v<0$ region is formed by the propagation of the slow shock produced in the process of magnetic reconnection along a uniform background field.
Furthermore, from the figures \ref{fig:slow_current}a--c, we can observe that the sheared loop is displaced in the negative y-axis direction with time, and with it, the region of $\mathrm{div}v<0$ is also displaced in the negative y-axis direction.
It can be understood that the upward displacement of the slow shock in the $y=\mathrm{const}$ cross section observed in figures \ref{fig:reco_rate}e-h corresponds to the negative y-axis displacement of the slow shock propagating through a uniform background magnetic field.
In fact, the slow shock rising speed in figure \ref{fig:reco_rate} is less than $0.1v_0$, which corresponds to the sheared loop speed, namely the reconnection inflow.

\section{Discussion}

\subsection{Formation mechanism of the jet-like structure}

\begin{figure*}[]
  \begin{center}
  \includegraphics[bb= 0 0 1150 909,width=160mm]{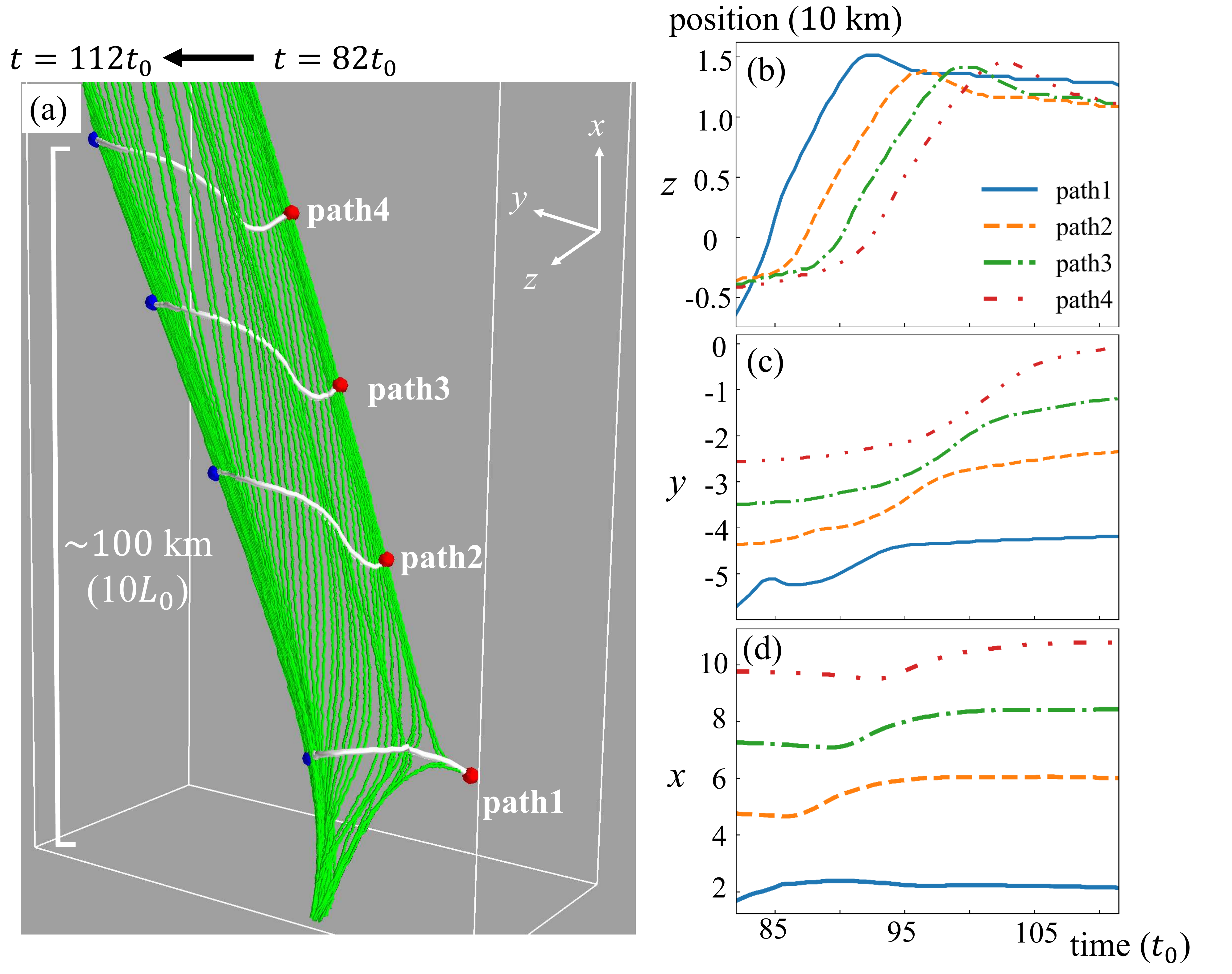}
  \end{center}
  \caption{%
 (a): 3D view of the paths traced by Lagrangian methods.
Red and blue points show the start and endpoints of time development.
White lines mean each path and green lines show magnetic field lines passing through path 2 at each time.
(b)-(d): Time development of z, y, and x components of each path.
The same type of lines indicates the same paths.
}%
  \label{fig:laglag}
\end{figure*}

\begin{figure}
  \begin{center}
  \includegraphics[bb= 0 0 413 1306,width=65mm]{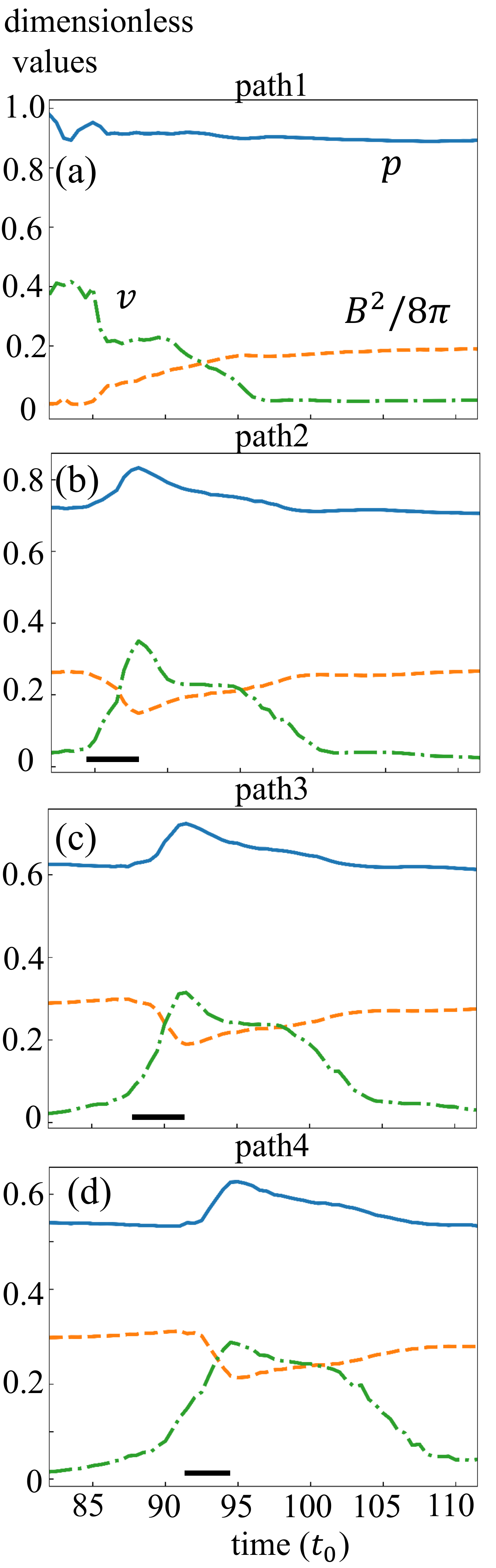}
  \end{center}
  \caption{%
 Time development of physical values in each path tracked by the Lagrangian method.
Solid blue, dashed orange, and dash-dot green lines mean gas pressure $p$, magnetic pressure $B^2/8\pi$, and plasma speed $|\bm{v}|$.
The units are $3.97\times10^4\,\mathrm{erg}\,\mathrm{cm}^{-3}$ for gas and magnetic pressure and $6.32\times10^5\,\mathrm{cm}\,\mathrm{s}^{-1}$ for plasma speed.
The black lines shown in (b), (c), and (d) indicate the time when the gas pressure increases in each path.
  }%
  \label{fig:lag_phy}
\end{figure}

\begin{figure*}[]
  \begin{center}
  \includegraphics[bb= 0 0 950 1278,width=150mm]{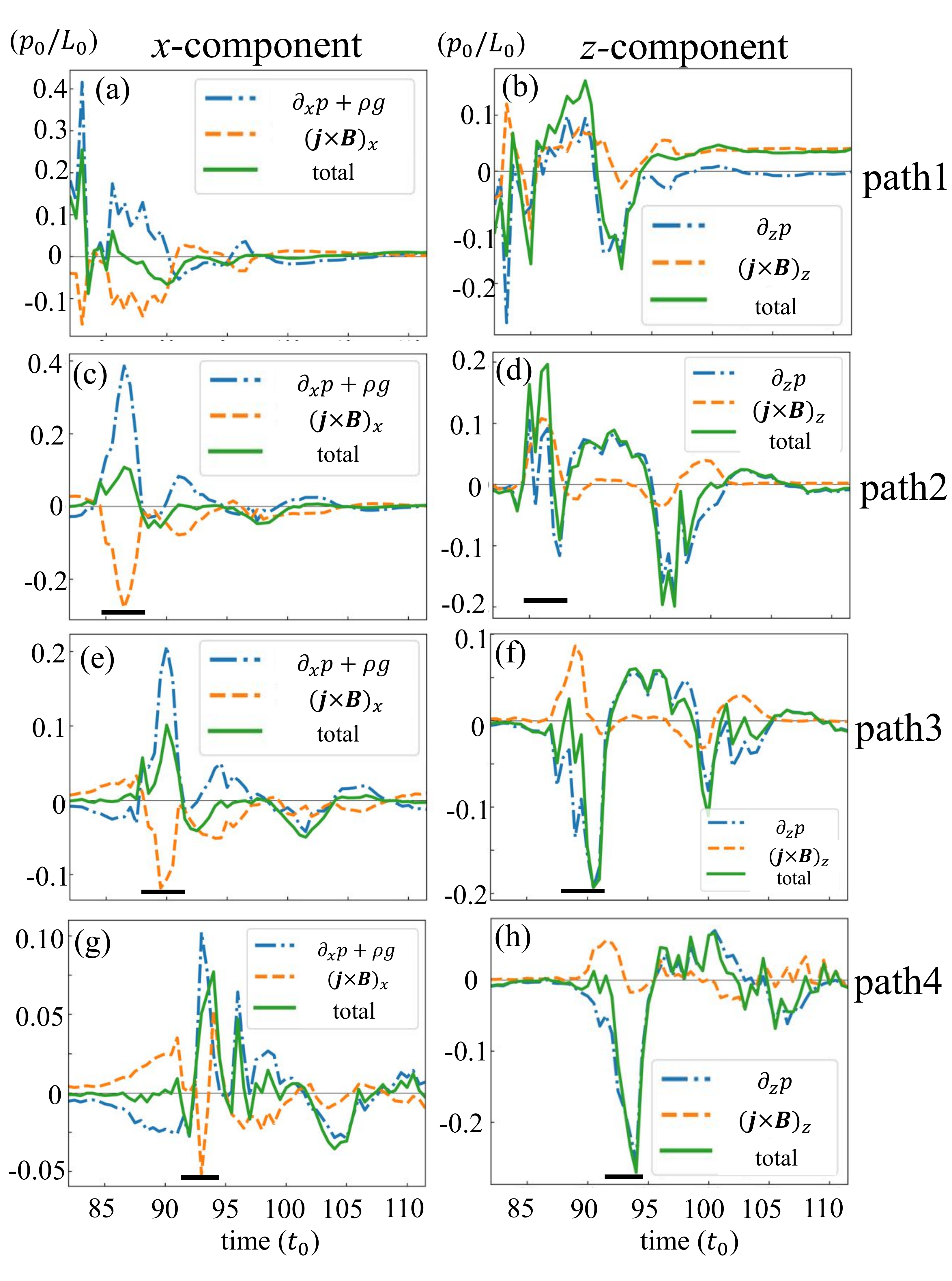}
  \end{center}
  \caption{%
 Time evolution of the forces acting on the plasma on each path.
  The left column shows the $x$-component of the force, and the right column shows the $z$-component.
  Each row shows the respective path shown in Figure \ref{fig:laglag}.
  Dash-dot blue, dashed orange, and solid green lines mean gas pressure gradient and gravity, Lorentz force, and total force.
  The black lines shown in (c)--(h) indicate the time when the gas pressure increases in each path in figure \ref{fig:lag_phy}.
}%
  \label{fig:lag_power}
\end{figure*}

     To determine the formation mechanism of the jet-like structure, 
we display in figure \ref{fig:laglag}a, the Lagrangian trajectories of typical fluid particles on the same reconnected magnetic field line.
For determining a reconnected field line, we select a point where the total force in the $x$-direction is greater than $0.1\rho_0v_0^2L_0^{-1}$ at $t=82t_0$.
Then, we select four points on the magnetic line passing through the point.
We assume the time interval $dt$ is $0.5t_0$.
In figures \ref{fig:laglag}b--d, we indicate the position $(x,y,z)$ of those fluid particles on these as a function of time.
From these figures, we can observe that the plasma rises; however, they do not move significantly when the magnetic tension attempts to straighten the reconnected field lines. 
In each path, the plasma particle is displaced only approximately $10\,\mathrm{km}$.
This indicates that the jet-like structure of approximately $100\,\mathrm{km}$ observable in figures \ref{fig:2d_rho} and \ref{fig:2d_jet_like} is not formed by plasma motion; rather,  it is formed by some wave propagation.

\begin{figure*}[]
  \begin{center}
  \includegraphics[bb= 0 0 587 1366 ,width=90mm]{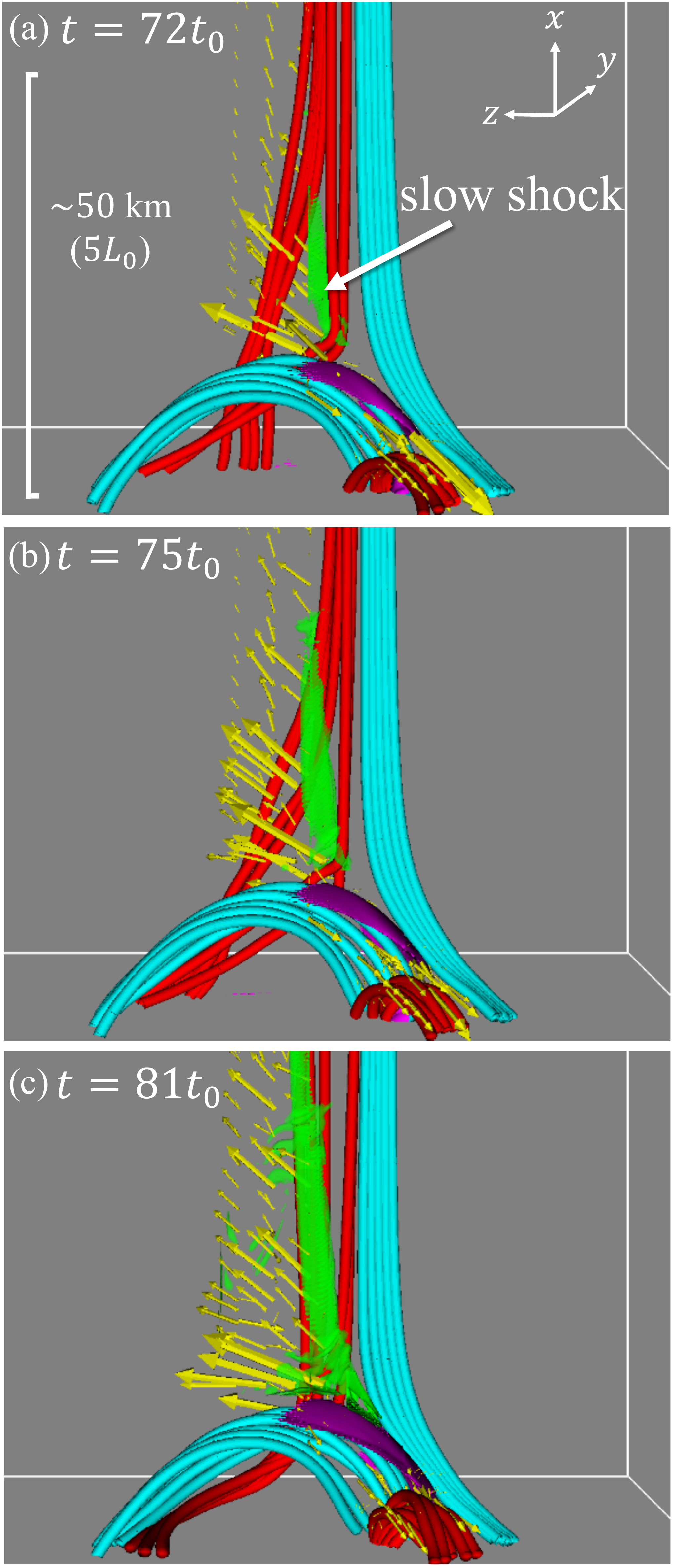}
  \end{center}
  \caption{%
  3D diagrams where the slow shock created by reconnection propagates in the direction of the background magnetic field.
  Green surface means $\mathrm{div} v=-0.09/t_0$ and $-0.07/t_0$, which corresponds to a slow shock created by reconnection.
  (a), (b), and (c) show $t=72t_0$, $75t_0$, and $81t_0$ cases.
  the color of magnetic field lines, the pink surface, and the yellow arrows indicate the same as figure \ref{fig:3d_jet}.
 }%
  \label{fig:3d_jet_time}
\end{figure*}

Figure \ref{fig:lag_phy} displays the time development of the gas pressure, magnetic pressure, and plasma speed along the trajectories indicated in figure \ref{fig:laglag}.
In figures \ref{fig:lag_phy}b, \ref{fig:lag_phy}c, and \ref{fig:lag_phy}d, we can observe that gas pressure increases and the magnetic pressure decreases when the plasma speed increases rapidly.
Furthermore, we can observe that this is the only time when the gas pressure increases in the time evolution of each path.
From these facts, it can be understood that the gas compression is caused by the slow shock and the jet-like structure is the result of the slow shock propagating along a uniform background magnetic field observable in figures \ref{fig:reco_rate}d, \ref{fig:reco_rate}h, \ref{fig:reco_rate}l, \ref{fig:slow_current}, and \ref{fig:3d_jet_time}.

Figure \ref{fig:lag_power} displays the temporal evolution of the forces acting on each path indicated in figure \ref{fig:laglag}.
We can observe that the upward gas pressure gradient is predominant in the $x$-component of the force when the gas pressure is increased.
Conversely, in the $z$-component, the positive Lorentz force functions mainly at first; however, the difference with the gas pressure gradient is small, and the negative gas pressure gradient ultimately prevails.
Considering that the magnetic field and gas pressure change rapidly in the slow shock and the value of plasma $\beta$ is approximately one, the behavior of these forces is consistent with the behavior when passing through the slow shock.
After passing through the slow shock, the force acting on the plasma becomes increasingly less and settles into a new equilibrium state.
These facts also indicate that the jet-like structure is not formed by reconnection outflow re-accelerated in the direction of the background magnetic field; rather, it is formed by compression due to the slow shock propagating in the background field's direction.
That is, this jet-like structure is not mass motion; rather, it is a slow shock propagation.

Note that this formation mechanism is different from previous studies regarding coronal and chromospheric cases.
In the coronal 2D simulation by \citet{1996PASJ...48..353Y}, hot jets are created by gas pressure increasing with a fast shock formed when reconnection outflow collides with the background field.
In the 2D simulation in both the corona and chromosphere \citep{1996PASJ...48..353Y,2013PASJ...65...62T}, the magnetic centrifugal force working in reconnected field lines drives cool jets.
In these cases,  a slow shock propagates approximately in the direction of the reconnection outflow, and the outflow is re-accelerated in the direction of the background field by the gas pressure gradient or Lorentz force.
These mechanisms are different from our photospheric case, where the slow shock propagates approximately along the reconnected field lines and the reconnection outflow is not re-accelerated.
3D simulation in the corona (e.g., \cite{2009ApJ...691...61P}) shows that jets are accelerated by nonlinear torsional Alfv\'en waves released from the twisted sheared loop.
In this case, the jets are accelerated by the magnetic pressure gradient.
In our simulation, a twist of the reconnected field lines can be observed (figure \ref {fig:3d_jet}) and Alfv\'en waves are generated (figure \ref{fig:energy_flux}); however, they do not function well regarding the compression and acceleration of the plasma. 

These differences from previous studies are due to our reconnection occurring in the region $\beta\sim3-4$.
In the $\beta\sim3-4$ case, the reconnection outflow is subsonic; hence, a fast shock cannot be created where the outflow collides with the background field.
Furthermore, because the magnetic energy is not dominant, it is difficult to accelerate the plasma using only the Lorentz force.
However, a slow shock can be made if reconnection occurs.

\begin{figure}
  \begin{center}
  \includegraphics[bb= 0 0 592 1277,width=60mm]{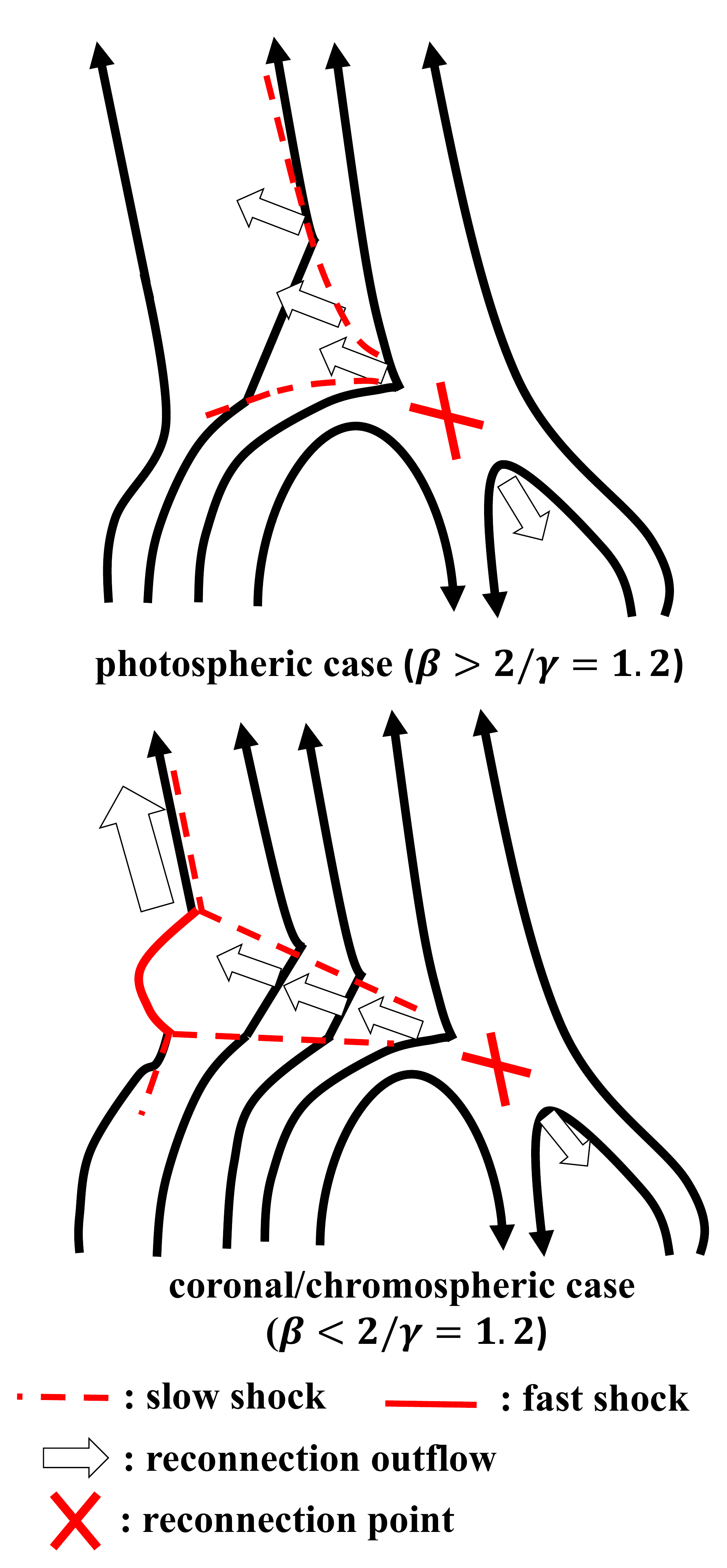}
  \end{center}
  \caption{%
 Schematic diagrams of the difference in slow shock propagation direction due to the difference in plasma $\beta$.
 $\gamma=5/3$ is the specific heat ratio.
 }%
  \label{fig:ponti}
\end{figure}

There are several possible causes for the slow shock immediately propagating in the background direction.
First, our reconnection outflow is subsonic; hence, it can bend before it collides with the background field.
Secondly, the phase speed of the slow shock in the direction of the magnetic fields is Alfv\'en speed when $\beta\gg1$.
Therefore, the slow shock is unlikely to propagate in the outflow direction, unlike in the case of $\beta\ll 1$.
Consequently, the slow shock is bent immediately after being formed on the current sheet and propagating in the background magnetic field (see figure \ref{fig:ponti}).

Note that if these photospheric jet-like phenomena are observed, it is expected that the apparent speed obtained from the imaging observation and the line-of-sight speed obtained from the spectroscopic observation are different.
In our simulation, the jet-like structure propagates at approximately $0.8v_0=5.04\,\mathrm{km}\,\mathrm{s}^{-1}$, corresponding to Alfv\'en speed.
Conversely, the speed of the plasma accelerated by the slow shock is approximately $0.35v_0=2.2\,\mathrm{km}\,\mathrm{s}^{-1}$, which is approximately half the propagation speed of the slow shock.

\subsection{MHD wave propagation toward upper atmosphere}

To determine the energy flux of Alfv\'en wave $F_\mathrm{A}$, slow mode wave $F_\mathrm{slow}$, and fast mode wave  $F_\mathrm{fast}$ are passing through the $x=\mathrm{const}$ plane, we calculate these as follows:
\begin{eqnarray}
\lefteqn{F_\mathrm{A} =   \frac{1}{4\pi}\frac{\int B_{0x}\bm{v}_{\perp}\cdot\bm{\delta B}_{\perp} dS}{\int dS}}    \label{eq:v_x}       \\
&&F_\mathrm{slow} =  \frac{\int\frac{B_{0x}}{B_0}v_{\parallel}\delta p dS}{\int dS}    \label{eq:vnorm} \\
&&F_\mathrm{fast} =  \frac{1}{4\pi}\frac{\int B_{0x}\delta B_{\parallel}v_{\perp} dS}{\int dS}    \label{eq:fast}\\
&&F_\mathrm{kin} =  \frac{\int \rho v^2v_{x} dS}{\int dS}.  \label{eq:kin}
\end{eqnarray}     
In the above equations, $\bm{\delta B} = \bm{B} - \bm{B}_0$ and $\delta p = p - p_0$.
The subscript “0”, $\perp$, and $\parallel$ are the values in the initial condition, direction normal, and parallel to the initial magnetic fields, respectively.
$F_{\mathrm{kin}}$ is the kinetic energy flux passing through the $x=\mathrm{const}$ plane.†
We set an integration range as the region where the absolute value of each energy flux exceeds 25$\%$ of the maximum value of the absolute value of the energy flux.
We measure these values in the $x=7.5L_0$, $15L_0$, and $22.5L_0$ planes, where plasma $\beta\simeq 2.0$, $1.25$, and $0.8$. 
Note that equations (\ref{eq:vnorm}) and (\ref{eq:fast}) correspond to magnetosonic waves in the region where $\beta$ is less than one, and do not strictly correspond to the wave's energy flux.
This is because the direction of the magnetosonic wave oscillation changes when $\beta$ is greater than one.
However, regardless of the  $\beta$ value, equations (\ref{eq:v_x}) and (\ref{eq:vnorm}) correspond approximately to the components of the Poynting and enthalpy fluxes in the direction of the initial magnetic field, and equation (\ref{eq:fast}) corresponds to the component normal to the initial field of the Poynting flux.

\begin{figure*}[]
  \begin{center}
  \includegraphics[bb= 0 0 852 980,width=160mm]{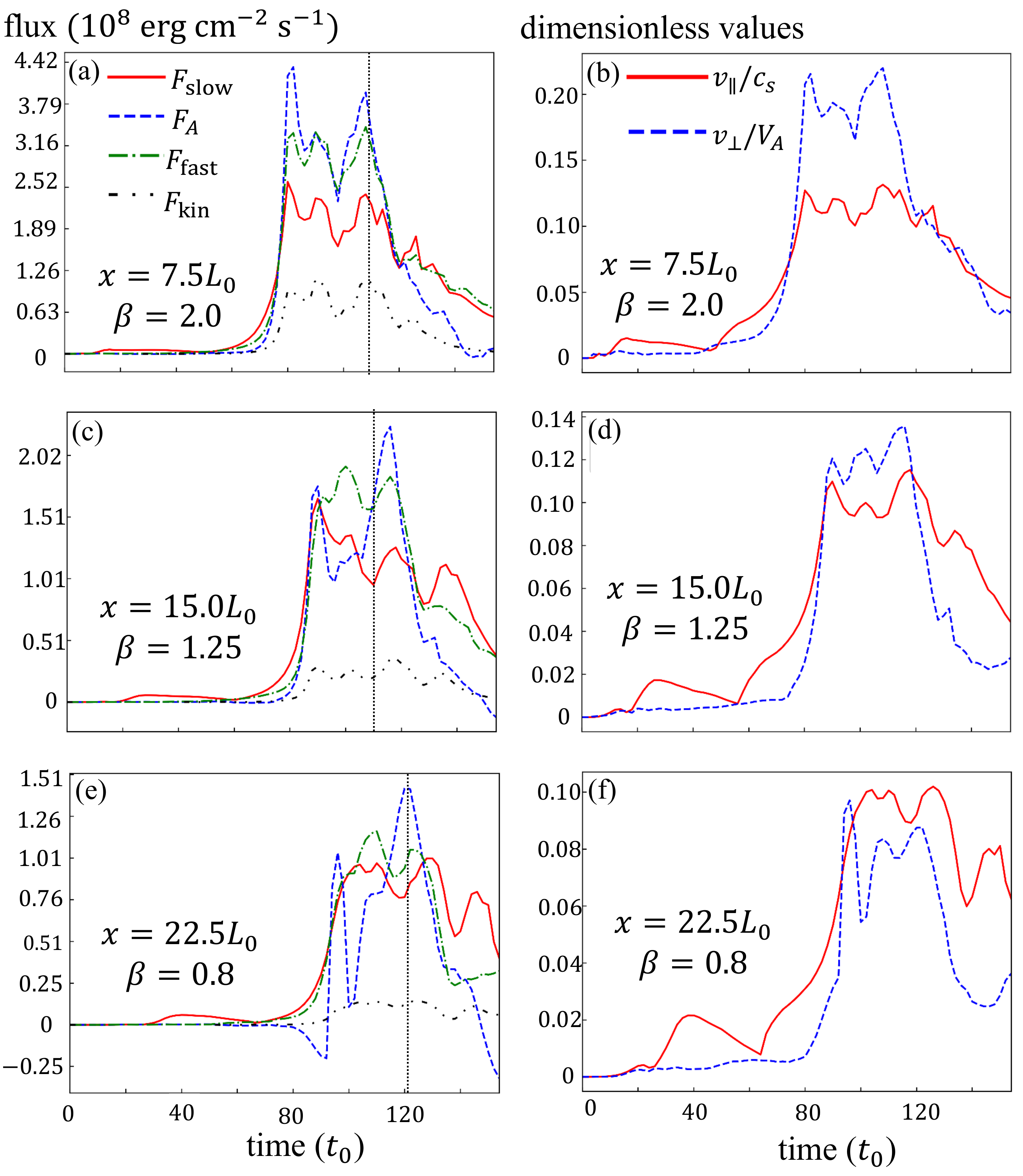}
  \end{center}
  \caption{%
 (a), (c), (e): Time development of MHD wave energy flux passing through $x=7.5L_0$, $15L_0$, and $22.5L_0$ plane.
Dashed blue, solid red, and dash-dot green lines show Alfv\'en, slow, and fast mode.
Dash-dot-dot black lines indicate kinematic energy flux.
The dotted line shown in (a) and (c) indicates $t=110t_0$, the time shown in figures \ref{fig:flux_plot_x7_5} and \ref{fig:flux_plot_x15}.
The dotted line indicated by (e) denotes $t=122t_0$, which is the time indicated in figure \ref{fig:flux_plot}.
(b), (d), (f): Time development of Mach number of Alfv\'en and sound wave.
Blue dashed and red solid lines show Alfv\'en and sound waves.
}%
  \label{fig:energy_flux}
\end{figure*}

Figures \ref{fig:energy_flux}a, \ref{fig:energy_flux}c, and \ref{fig:energy_flux}e display the result.
At first, all mode waves are released, which corresponds to the sheared loop rising phase (figure \ref{fig:2d_rho}b).
The release of all modes in reconnection is the same as the results of \citet{2010PASJ...62..993K}.
The reason for Alfv\'en wave dominance at $x=7.5L_0$ is that the velocity in the $z$-direction, which is perpendicular to the initial background magnetic field, is the greatest.
Subsequently, the twist of the sheared loop is released by magnetic reconnection, and the energy flux of the Alfv\'en mode becomes dominant in all planes.
These features indicate that the jet-like structure is in Alfv\'enic motion, and also suggest that the reconnection process is similar to previous studies of coronal jets  (e.g., \cite{2009ApJ...691...61P,2013ApJ...769L..21A,2018ApJ...852...98W}).
In each plane, all modes have approximately $10^8\,\mathrm{erg}\,\mathrm{cm}^{-2}\,\mathrm{s}^{-1}$ in the $x=22.5L_0$ plane.
Moreover, the kinetic energy flux is smaller than the energy flux of the waves.
This indicates that the nonlinearity of the plasma's motion forming the jet-like structure is not strong.

Figures \ref{fig:energy_flux}b, \ref{fig:energy_flux}d, and \ref{fig:energy_flux}f  display the mean value of the Mach number $M_s=v_{\parallel}/c_s$ and Alfv\'en Mach number $M_A=v_{\perp}/V_A$ in the area where slow and Alfv\'en mode is passing at each time.
From these figures, we can observe that the peak value of each Mach number is approximately 0.1; this value is consistent with the case of the coronal jets (\cite{1999AIPC..471...61Y}).
Furthermore, the Alfv\'en Mach number decreases with height, yet the Mach number of sound waves remains virtually unchanged.
This is because, in our numerical settings, the  Alfv\'en speed increases with height and the sound speed is uniform.

\begin{figure*}[]
  \begin{center}
  \includegraphics[bb= 0 0 1134 710,width=160mm]{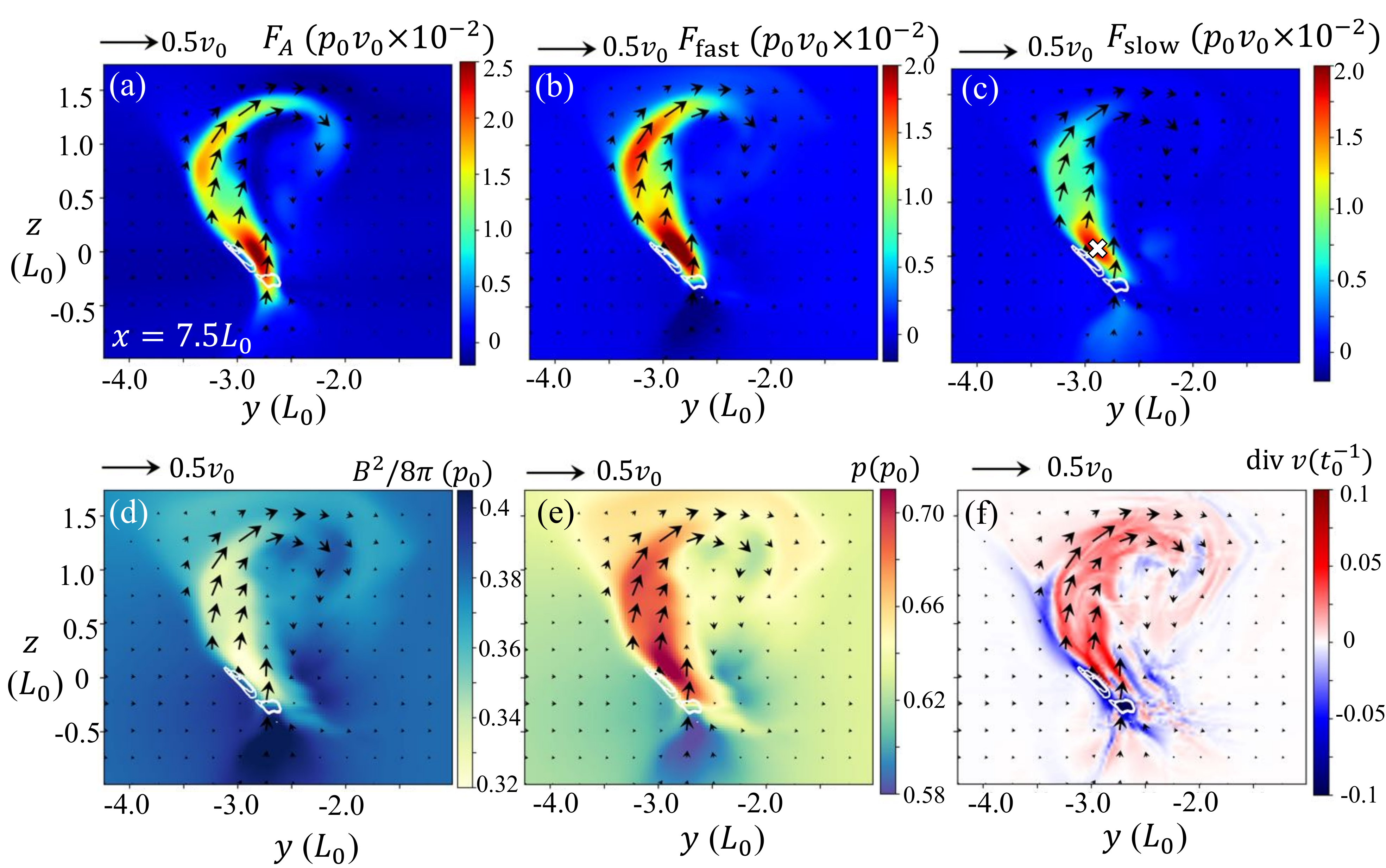}
  \end{center}
  \caption{%
  Energy flux of each wave and the distribution of magnetic and gas pressures and $\mathrm{div}v$ in the $x=7.5L_0$ plane at $t=110t_0$.
  Black arrows show velocity  in the plane.
  White contours indicate $\mathrm{div}v=-0.07t_0^{-1}$ and $-0.09t_0^{-1}$.
  The white x mark in (c) indicates the area where the magnetic field line is passing through in figure \ref{fig:flux_3d}.
}%
  \label{fig:flux_plot_x7_5}
\end{figure*}

\begin{figure*}[]
  \begin{center}
  \includegraphics[bb= 0 0 1134 738,width=160mm]{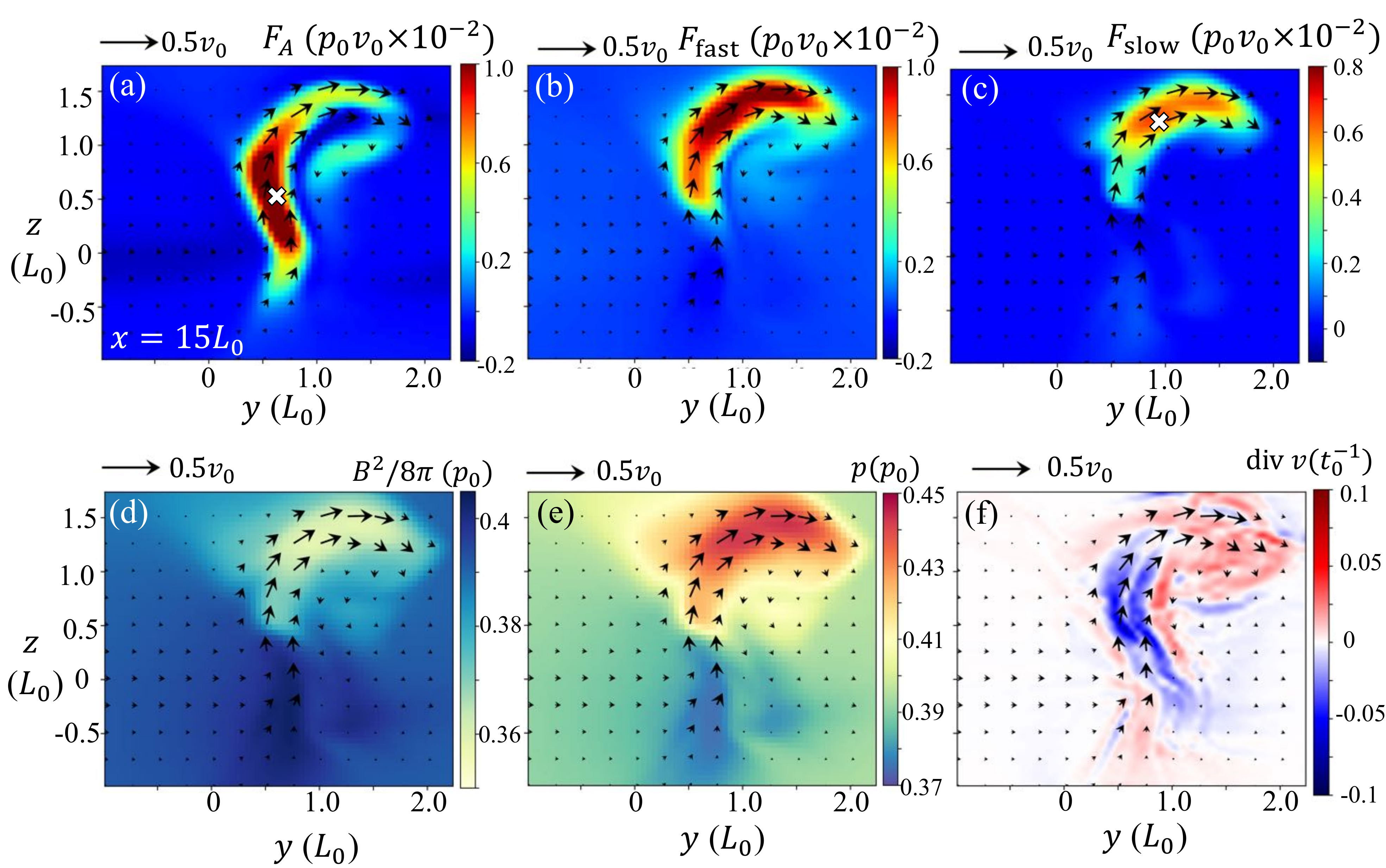}
  \end{center}
  \caption{%
  Energy flux of each wave and the distribution of magnetic and gas pressures and $\mathrm{div}v$ in the $x=15L_0$ plane at $t=110t_0$.
  Black arrows show velocity  in the plane.
   The white x mark in (c) indicates the area where the magnetic field line is passing through in figure \ref{fig:flux_3d}.
}%
  \label{fig:flux_plot_x15}
\end{figure*}

\begin{figure*}[]
  \begin{center}
  \includegraphics[bb= 0 0 1134 738,width=160mm]{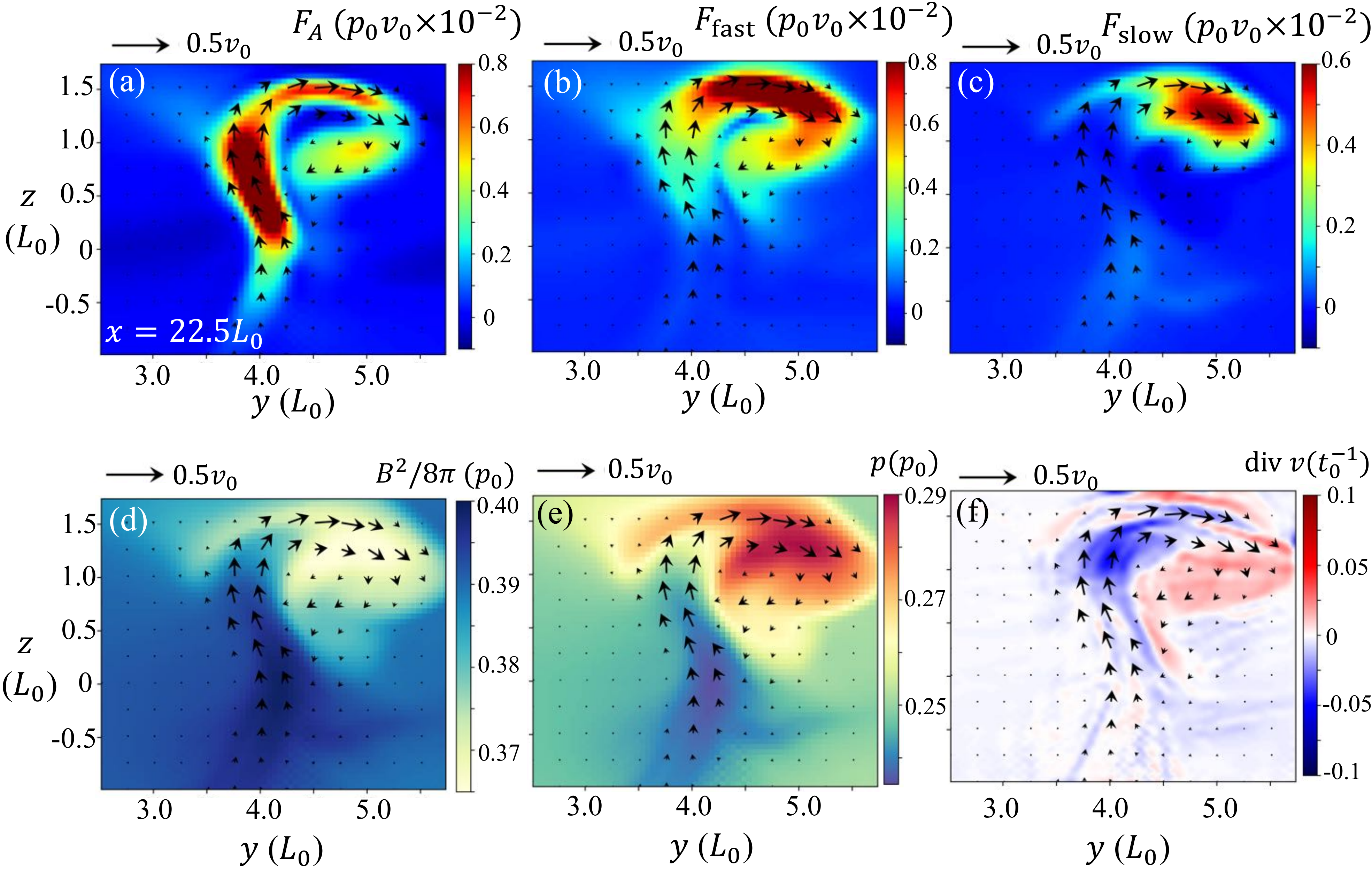}
  \end{center}
  \caption{%
  Energy flux of each wave and the distribution of magnetic and gas pressures and $\mathrm{div}v$ in the $x=22.5L_0$ plane at $t=122t_0$.
  Black arrows show velocity  in the plane.
}%
  \label{fig:flux_plot}
\end{figure*}

Figures \ref{fig:flux_plot_x7_5},  \ref{fig:flux_plot_x15}, and \ref{fig:flux_plot} display the energy flux of each wave and the distribution of the magnetic and gas pressures and $\mathrm{div}v$ in the $x=7.5L_0$ and  $x=15L_0$ planes at $t=110t_0$ and $x=22.5L_0$ plane at $t=122t_0$.
From these figures, we can observe that the gas pressure increases and magnetic pressure decreases in the region of intense slow mode energy flux.
Moreover, the jet-like structure can be observed to be in vortex motion.
In figure \ref{fig:flux_plot_x7_5}, the region with strong wave energy flux is in virtually the same place, adjacent to the region with strong $\mathrm{div}v<0$.
In figures \ref{fig:flux_plot_x15} and \ref{fig:flux_plot}, the regions with strong wave fluxes differ from each mode.
In particular, it can be observed that the strong Alfv\'en wave region is where the magnetic pressure is strong, which is different from the region of the strong slow mode.
Moreover, the strength of $\mathrm{div}v$ is weaker than in the $x=7.5L_0$ plane.

\begin{figure*}[]
  \begin{center}
  \includegraphics[bb= 0 0 777 512,width=160mm]{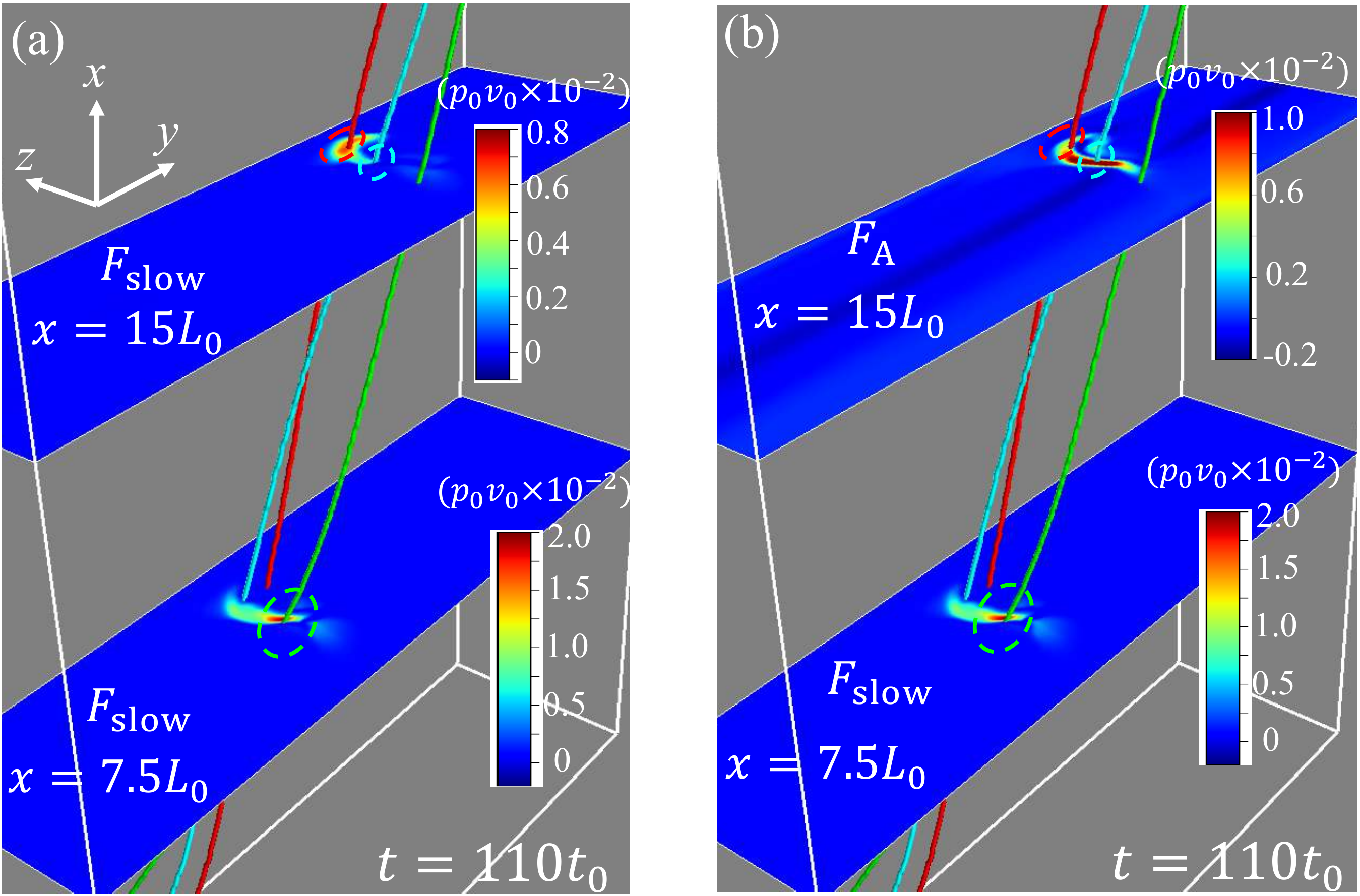}
  \end{center}
  \caption{%
  3D relationship between the $yz$ plane showing the distribution of energy fluxes and magnetic field lines at $t=110t_0$.
  In (a), we show the distribution of slow mode energy flux in the $x=15L_0$ plane shown in figure \ref{fig:flux_plot_x15}c, and in (b), we show that of Alfv\'en mode shown in figure \ref{fig:flux_plot_x15}a.
  In both figures, the $x=7.5L_0$ plane shows the distribution of slow mode energy flux shown in figure \ref{fig:flux_plot_x7_5}c.
  The green line indicates a magnetic field line passing through a region with a strong energy flux in the $x=7.5L_0$ plane.
  The light blue line shows a magnetic field line passing through a region with a strong energy flux of Alfv\'en mode in the $x=15L_0$ plane, and the red linesshow a magnetic field line passing through a region with a strong energy flux of slow mode in the same plane.
 The three dashed lines surrounding the magnetic field lines indicate the region through which each magnetic field line passes.
}%
  \label{fig:flux_3d}
\end{figure*}

\begin{figure}
  \begin{center}
  \includegraphics[bb= 0 0 437 1305,width=60mm]{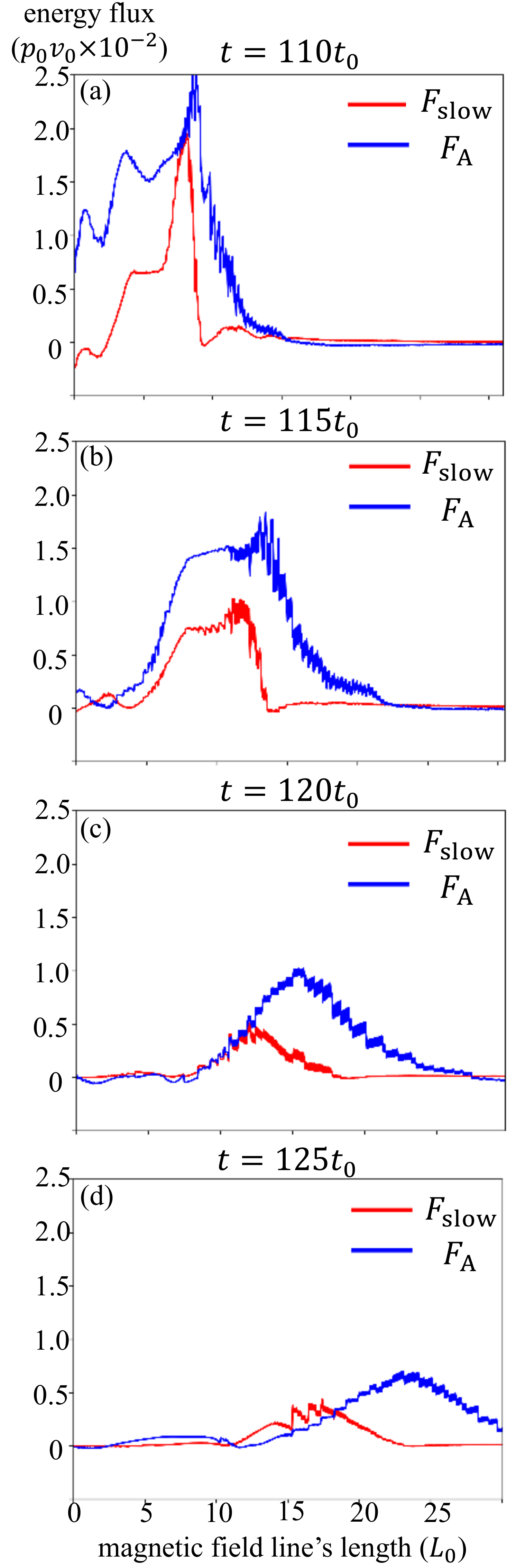}
  \end{center}
  \caption{%
  Time evolution of the distribution of energy fluxes along the green magnetic field line shown in figure \ref{fig:flux_3d}, which is passing through a region with a strong slow mode in the $x=7.5L_0$ plane.
The time evolution is calculated by tracking the magnetic field line in a Lagrangian way.
The horizontal axis shows the length of the magnetic field line when the bottom surface is set to zero.
The red and blue lines show the slow and Alfv\'en  mode energy fluxes, respectively.
(a), (b), (c), and (d) show the distribution at $t = 110t_0$, $115t_0$, $120t_0$, and $125t_0$, respectively.
 }%
  \label{fig:flux_propai}
\end{figure}

The reason the position of the peak of each mode is shifted as it rises is that the propagation speed of the slow mode and the Alfv\'en mode is different in the upper region.
Figure  \ref{fig:flux_3d} displays  the 3D relationship between the $yz$ planes indicateing the distribution of the energy fluxes and magnetic field lines at $t = 110t_0$.
Considering the vortex motion observable in figures \ref{fig:flux_plot_x7_5}, \ref{fig:flux_plot_x15}, and \ref{fig:flux_plot}, we can observe that a magnetic field line, along which waves propagate, moves in the order of the green, light blue, and red line with time.
Figure \ref{fig:flux_propai} displays the time evolution of the distribution of energy fluxes along the magnetic field lines passing through a region with a strong slow mode in the $x = 7.5L_0$ plane.
From this figure, we can observe that the positions of the peaks of the Alfv\'en mode and slow modes are virtually the same in the lower part, and that the Alfv\'en mode precedes in slow mode in the upper part.
This is because where $\beta$ is greater than $2/\gamma$, the slow mode propagates at the Alfv\'en velocity along the magnetic field line; however, below $2/\gamma$, it propagates at the sound speed.
From these facts, the position of the energy flux peaks appearing in the $yz$ cross section is different in the upper part because the slow mode passes after the Alfve\'n mode along the vortex-moving magnetic field lines.

We also investigated the effects of MHD waves generated from photospheric anemone jet-like structures on the upper atmosphere.
From a simple estimate, we determined that the MHD wave's energy flux achieving a height of $500\,\mathrm{km}$ was approximately $4\times10^7\,\mathrm{erg}\,\mathrm{cm}^{-2}\,\mathrm{s}^{-1}$ (see Appendix 1).
This value is marginally greater than the amount required for chromospheric heating  (\cite{1977ARA&A..15..363W}).
Because of the small lifetime of the phenomenon, these waves are high frequency, and we can observe from figure \ref{fig:energy_flux} that the period is approximately $20t_0\sim30\,\mathrm{s}$.
This implies that the photospheric anemone jet-like structure could be one of the origins of the high-frequency Alfv\'en waves that have been observed in the spicules \citep{2009A&A...497..525H,2011ApJ...736L..24O}, though other candidates exist, such as mode conversions of longitudinal to transverse waves \citep{2018ApJ...854....9S}.
Furthermore, we extended the computational domain to the corona and performed 1D hydrodynamic simulations.
From the results of the simulations, we determined that the photospheric jet-like structure can also influence the spicule formation (see Appendix 2).
Note that these estimates are simplistic and should be studied in more detail in future papers.

\section{Conclusion}
We performed 3D MHD simulation of anemone jet-like structures in the solar photospheric parameter.
From the results of the simulation, a jet-like structure was induced by magnetic reconnection.
The length, width, lifetime, and apparent velocity of the jet-like structure were extensions of the coronal and chromospheric anemone jets.
This jet-like structure was formed by the propagation of the slow shock generated by magnetic reconnection.
These facts indicate that the anemone jet-like structure, which can be explained by the unified model \citep{1999Ap&SS.264..129S,2007Sci...318.1591S}, is expected to exist in the solar photosphere; however, the formation process of the jet-like structure is different from that in the low $\beta$ environment of the chromosphere and corona.
In the present study, we simulated a jet-like structure with a length of approximately $100\,\mathrm{km}$.
However, such a jet-like structure is expected to be formed in an environment where the plasma $\beta$ is greater than one, even though the scales are different.

We also confirmed, for the first time, that the magnetosonic and Alfv\'en waves are generated by magnetic reconnection in the solar photospheric parameters in a 3D manner .
These waves are high frequency, and their non-dimensional amplitudes are comparable with those of the coronal case.
Furthermore, the MHD wave energy fluxes were generated to the degree that could influence the local chromospheric heating and the formation of spicules.

\begin{ack}
We would like to Mr. T. Sakaue for his useful comments regarding the proposed numerical simulation and interpreting the simulation results.
We also thank Dr. T. Yokoyama and Dr. S. Toriumi for their fruitful discussions.
We are deeply grateful to Dr. K. Tomida for instructing us on the use of Athena++.
Furthermore, we wish to thank the anonymous referee for helpful comments that led to improvements in this work.
We would like to thank Editage (www.editage.com) for English language editing.
Numerical computations were carried out on Cray XC50 at Center for Computational Astrophysics, National Astronomical Observatory of Japan.
\end{ack}

\section*{Supporting Information}
The following Supporting Information is available in the online version of this article. E-movies 1 and 2.

\appendix

\section{Effect of chromospheric heating}

Figure  \ref{fig:energy_flux}e indicates that each mode has approximately $10^8\,\mathrm{erg}\,\mathrm{cm}^{-2}\,\mathrm{s}^{-1}$ in the $x=22.5L_0=225\,\mathrm{km}$ plane.
To discuss the effect of chromospheric heating, we estimate the amount of energy flux that can reach the chromosphere at a height of $\sim500\,\mathrm{km}$ as follows.
We consider an Alfv\'en wave and slow mode wave propagating along a vertical magnetic flux tube with cross section $S$.
First, we assume a pressure balance between the inside and outside flux tube, $ B^2/8\pi \sim p\propto e^{-x/H}$; then, $B\propto e^{-x/2H}$.
Secondly, we assume a magnetic flux conservation $BS=\mathrm{const}$; then, $S\propto e^{x/2H}$.
Finally, we assume energy conservation $FS=\mathrm{const}$, where $F$ is the energy flux.
Considering that a wave flux passing through the $x=225\,\mathrm{km}$ plane $F_{x=225\,\mathrm{km}}$ is approximately $10^8\,\mathrm{erg}\,\mathrm{cm}^{-2}\,\mathrm{s}^{-1}$, then a wave flux passing through the $x=500\,\mathrm{km}$ plane $F_{x=500\,\mathrm{km}}$ can be estimated.
\begin{eqnarray}
F_{x=500\,\mathrm{km}}\sim F_{x=225\,\mathrm{km}}\times e^{(225\,\mathrm{km}-500\,\mathrm{km})/2H}\nonumber  \\
\sim 4\times10^7\,\mathrm{erg}\,\mathrm{cm}^{-2}\,\mathrm{s}^{-1}  \label{eq:chro_flux}  
\end{eqnarray}
This value is approximately the energy flux required for heating the chromosphere in the active region $1.5\times10^7\,\mathrm{erg}\,\mathrm{cm}^{-2}\,\mathrm{s}^{-1}$ and greater than $4\times10^6\,\mathrm{erg}\,\mathrm{cm}^{-2}\,\mathrm{s}^{-1}$ in the quiet region (\cite{1977ARA&A..15..363W}).
Because these waves are high frequency, the transverse waves are easily converted to longitudinal waves by mode conversion, and the majority of these are dissipated in the chromosphere \citep{2010ApJ...710.1857M}.
Assuming the loop-like geometry found in the active regions, \citet{2010ApJ...712..494A} found that active regions may not be heated by Alfv\'en waves, based on the expansion factor of the loop and other observed facts.
Therefore, the waves released from the photospheric anemone jet-like structure are expected to contribute mainly to the heating of the chromosphere.
Note that we must consider the frequency of the jet-like structure to discuss the contribution to global chromospheric heating because these energy fluxes originate from a photospheric jet-like structure.

\section{1D simulation for the spicule formation}

To investigate how slow mode waves generated from the photospheric anemone jet-like structures propagate to the upper atmosphere along a vertical magnetic flux tube, we perform 1D non-magnetic hydrodynamic simulations.

For the numerical simulation, we use Athena++ code with the van Leer predictor-corrector scheme and Piecewise Linear Method \citep{2020arXiv200506651S}.
We solve the compressive hydrodynamic equation including uniform gravity.
The basic equations are as follows.
\begin{eqnarray}
\lefteqn{\frac{\partial \rho}{\partial t} + \frac{\partial}{\partial s}(\rho v_{\parallel})= 0 }      \label{eq:mass_1}      \\
&&\frac{\partial \rho v_{\parallel}}{\partial t} + \frac{\partial}{\partial s}(\rho v_{\parallel}^2) = -\frac{\partial}{\partial s}p +\rho g_{\parallel}      \label{eq: eom_1}   \\
&&\frac{\partial }{\partial t}(e + \frac{1}{2}\rho v_{\parallel}^2 ) + \frac{\partial}{\partial s} \bigl[(h + \frac{1}{2}\rho v_{\parallel}^2 )v_{\parallel} 
 \bigr] \nonumber
=\rho g_{\parallel}v_{\parallel}    \label{eq: energy_1}    \\
&&p = \frac{\rho R T}{\mu} \label{eq: eos_1}
\end{eqnarray}  
, where $e$ is the internal fluid energy and $h=p+e$  is the enthalpy.

For the coordinates of the 1D calculation, a parallel straight line is taken in the background field of the 3D calculation.
We set the height of the upper boundary as $12,000\,\mathrm{km}$ from the bottom of the photosphere.
The number of the mesh is $24,000$, and the grid spacing is uniform.
The normalization of the numerical calculation is performed in the same manner as the 3D calculation.
We use only the background field component for the gravitational acceleration.

As the initial condition, we assume hydrostatic equilibrium with initial temperature $T_{\mathrm{ini}}$.  
\begin{equation}
T_{\mathrm{ini}} = T_{\mathrm{pho}} + \frac{1}{2}(T_{\mathrm{cor}} - T_{\mathrm{pho}})(1 + \tanh(\frac{x - x_{\mathrm{tr}}}{w_{\mathrm{tr}}})) \label{eq:tempera}
\end{equation}
$T_{\mathrm{pho}} ,T_{\mathrm{cor}},x_{\mathrm{tr}}$, and $w_{\mathrm{tr}}$ are the temperature of the photosphere and corona, and the height and the thickness of the transition layer.
We set $T_{\mathrm{pho}}= 6000\,\mathrm{K} , w_{\mathrm{tr}} = 80\,\mathrm{km} $.
Then, along a reconnected field line in 3D simulation, we take out the thermodynamic quantity and velocity components parallel to the field and consider them as a perturbation.
For the parameters $T_{\mathrm{cor}}$ and $x_{\mathrm{tr}}$, we perform two cases, $T_{\mathrm{cor}} =170T_{\mathrm{pho}}=1.02\times10^6\,\mathrm{K}, x_{\mathrm{tr}} = 2300\,\mathrm{km}$ (quiet region case) and $T_{\mathrm{cor}}=400T_{\mathrm{pho}}=2.4\times10^6\,\mathrm{K},x_{\mathrm{tr}} = 1850\,\mathrm{km}$ (active region case).
The coronal pressure in the initial condition is a approximate agreement with the values reported in the observations (quiet region: \cite{2010ApJ...719..131I}, active region: \cite{2011ApJ...740....2W}).

For the boundary condition, we set the reflected boundary at the bottom and open boundary at the top. 
Note that in order to maintain the hydrostatic pressure equilibrium at the upper boundary,  we set the gravitational acceleration to zero smoothly above the height of $11,000\,\mathrm{km}$.
This height is sufficiently greater than that at which the contact discontinuity surface obtained from the calculation results rises, and thus this assumption does not substantially influence the calculation.

\begin{figure*}[]
  \begin{center}
  \includegraphics[bb= 0 0 836 397,width=160mm]{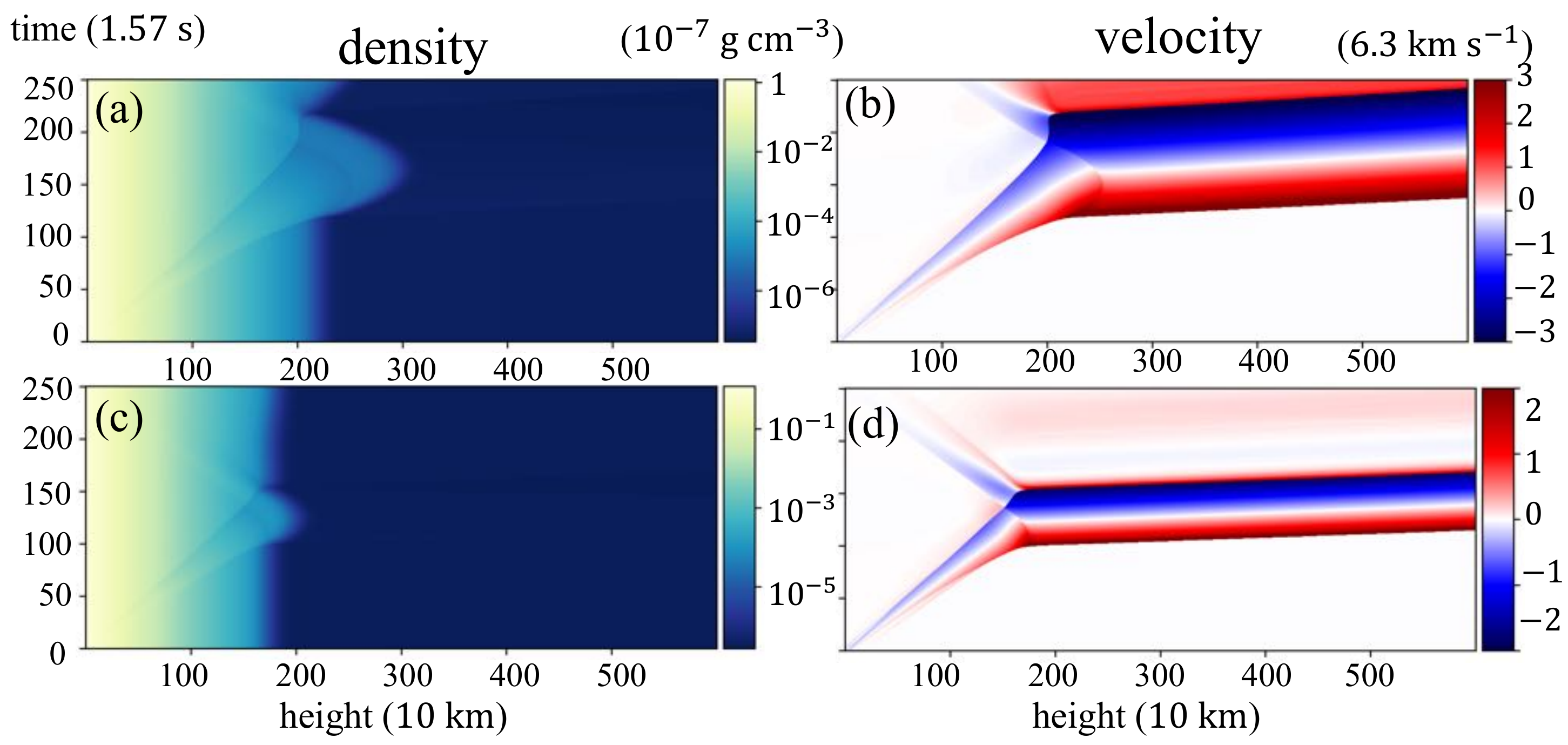}
  \end{center}
  \caption{%
 Results of 1D simulation.
 (a), (c): time-height plot of density.
 (b), (d): time-height plot of velocity.
 (a), (b) and (c), (d) correspond to quiet region and active region cases.
 }%
  \label{fig:3djet_zx}
\end{figure*}

Figure \ref{fig:3djet_zx} displays the results.
We can observe that the contact discontinuity is launched by shocks.
Figures \ref{fig:3djet_zx}a and \ref{fig:3djet_zx}b display the quiet region case, and we can observe that the maximum height is approximately $3200\,\mathrm{km}$, the maximum velocity is approximately $20\,\mathrm{km}\,\mathrm{s}^{-1}$, and the lifetime is approximately $160\,\mathrm{s}$.
\citet{2012ApJ...759...18P} and \citet{2012ApJ...750...16Z} performed a statistical study of spicules, and many of their results are consistent with our simulation.
Figures \ref{fig:3djet_zx}c and \ref{fig:3djet_zx}d display the active region case.
We can observe that the maximum height is approximately $2100\,\mathrm{km}$ (the maximum length is approximately  $300\,\mathrm{km}$), the maximum velocity is approximately $15\,\mathrm{km}\,\mathrm{s}^{-1}$, and the lifetime is approximately $100\,\mathrm{s}$.
\citet{2007ApJ...655..624D} and \citet{2010PASJ...62..871A} performed statistical studies of dynamic fibrils.
Their studies are consistent with our results regarding the maximum length and the maximum velocity.
The maximum length  and lifetime in our results are marginally shorter than their study, yet reasonably consistent with theirs.
These results suggest that photospheric anemone jet-like structures can be one of the origins of spicules and dynamic fibrils.

\begin{figure}
  \begin{center}
  \includegraphics[bb= 0 0 408 327,width=80mm]{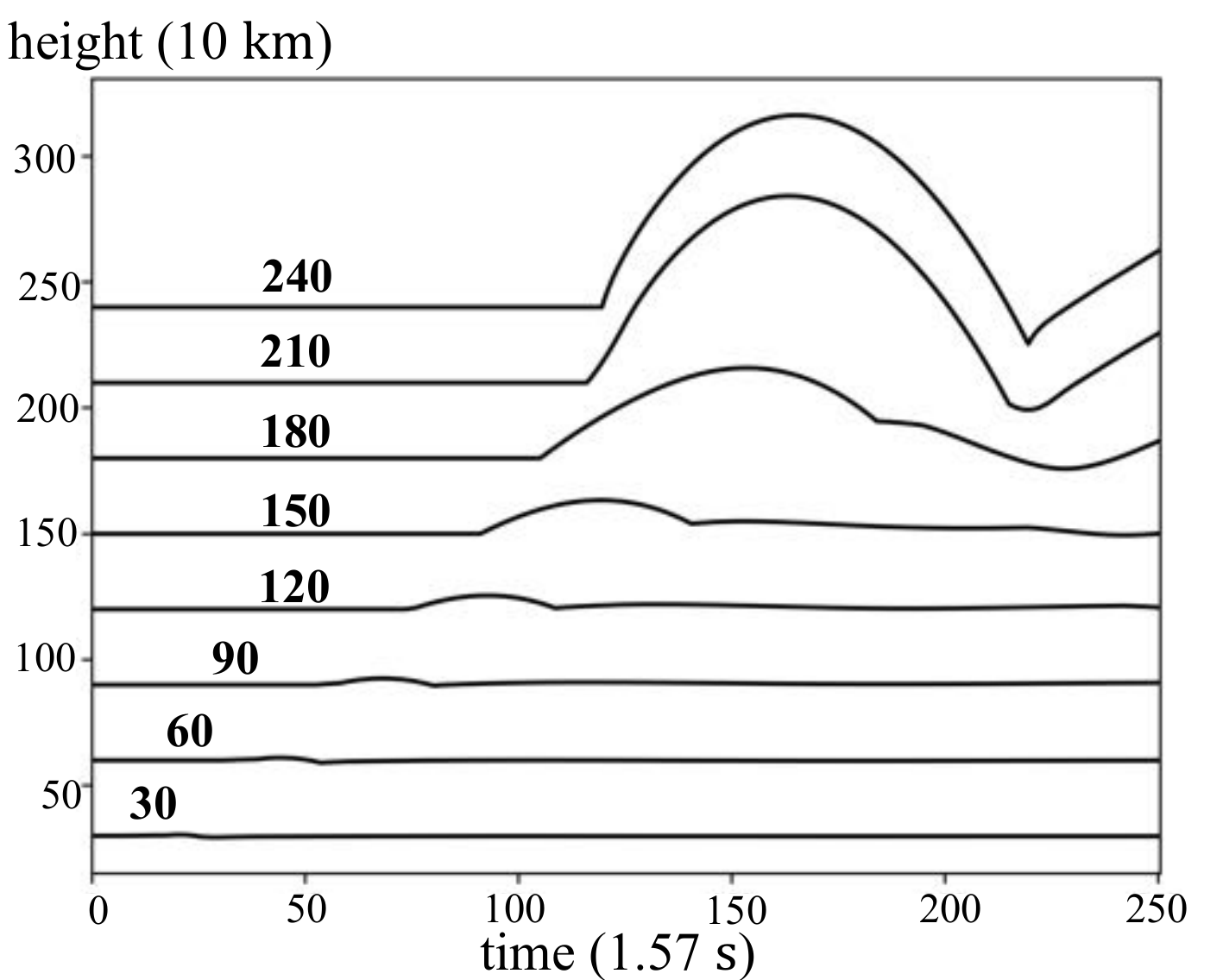}
  \end{center}
  \caption{%
Trajectories of the fluid particles in the case of the quiet region.
 The number above each curve indicates the initial position of each fluid particle.
  }%
  \label{fig:1d_coord}
\end{figure}

Figure \ref{fig:1d_coord} displays the trajectories of the fluid particles.
From this figure, we can observe that the fluid particles do not move to a great degree in the lower layer, yet near the transition region, they are significantly launched by the shock to form a jet.
Note that the behavior of the fluid particles in the lower layers in this figure is similar to that of those observable in figure \ref{fig:laglag}.
This is consistent with the result that the photospheric anemone jet-like structure is not a plasma flow.

\end{document}